\documentstyle[12pt]{article}
\newcommand{\NPB}[3]{{\it Nucl.~Phys.}~{\bf B#1},~#2~(19#3)}
\newcommand{\PLB}[3]{{\it Phys.~Lett.}~{\bf B#1},~#2~(19#3)}

\newcommand{\PRD}[3]{{\it Phys.~Rev.}~{\bf D#1},~#2~(19#3)}
\newcommand{\ZPC}[3]{{\it Z.~Phys.}~{\bf C#1},~#2~(19#3)}
\newcommand{\CPC}[3]{{\it Comput.~Phys.~Commun.}~{\bf #1},~#2~(19#3)}
\newcommand{\lesssim}{\stackrel{\scriptscriptstyle<}{\scriptscriptstyle\sim}}
\newcommand{\grsim}{\stackrel{\scriptscriptstyle>}{\scriptscriptstyle\sim}}
\newcommand{\qgen}{q}

\def\beq{\begin{equation}}
\def\enq{\end{equation}}
\def\ra{\rightarrow}

\def\D0{D\O}
\def\ETslash{\not{\hbox{\kern-4pt $E_T$}}}
\def\doublespaced{\baselineskip=\normalbaselineskip\multiply
    \baselineskip by 150\divide\baselineskip by 100}
\doublespaced
\begin{document}

\input psfig 
\baselineskip=\normalbaselineskip\multiply
\baselineskip by 150\divide\baselineskip by 100 \pagenumbering{arabic} 
\begin{titlepage}
{\small
\noindent
\hfill {ANL--HEP--PR--97--84}  \\
\rightline{CTEQ--710}\\
\rightline{MSUHEP--71027}
}
\vspace{0.4cm}
\begin{center} 
\large
{\bf Photon Pair Production With \\ Soft Gluon Resummation \\
     In Hadronic Interactions}
\end{center}
\vspace{0.4cm}
\begin{center}
{\bf C. Bal\'azs,$^{(a)}$
E.L. Berger,$^{(b)}$
S. Mrenna,$^{(b)}$ 
and~~C.--P. Yuan$^{(a)}$}
\end{center}
\vspace{0.2cm}
\begin{center}
{(a) Department of Physics and Astronomy,
Michigan State University \\
East Lansing, MI 48824, U.S.A. \\
(b) High Energy Physics Division,
Argonne National Laboratory\\
Argonne, IL 60439, U.S.A.
}
\end{center} 

\vspace{0.4cm}
\raggedbottom
\setcounter{page}{1}
\relax

\begin{abstract}
\noindent
The production rate and
kinematic distributions of isolated photon pairs produced 
in hadron interactions are studied.  The effects of the initial--state 
multiple soft--gluon emission to the scattering subprocesses 
$q {\bar q}, qg,$ and $gg \ra \gamma \gamma X$ 
are resummed with the Collins--Soper--Sterman 
soft gluon resummation formalism.
The effects of fragmentation photons 
from $qg\to \gamma q$, followed by $q\to \gamma X$,
are also studied.
The results are compared with data
from the Fermilab Tevatron collider.
A prediction of the production rate and kinematic distributions
of the diphoton pair in proton--nucleon reactions
is also presented.
\end{abstract}


\end{titlepage}
\newpage

\section{Introduction}

\indent\indent

An increasing amount of prompt diphoton data is becoming available from the
Tevatron collider and the fixed--target experiments at Fermilab.
A comparison of
the data to the calculation of the diphoton production rate and kinematic
distributions provides a test of many aspects of 
perturbative quantum chromodynamics (pQCD).  Furthermore, understanding the
diphoton data
is important for new physics searches. For example, diphoton production
is an irreducible background to the light Higgs boson decay mode 
$h\to \gamma\gamma $. The next--to--leading order (NLO) 
cross section for the $p\bar{p}\to\gamma\gamma X$ process
\cite{nloqq} was shown to describe well
the invariant mass distribution of the diphoton pair
after
the leading order (LO) $gg\to \gamma \gamma $
contribution (from one--loop box diagrams) was included \cite{box,nlo}. 
However, to
accurately describe the distribution of the transverse momentum of the
photon pair and the kinematical correlation of the two photons, a
calculation has to be performed 
that includes the effects of initial--state multiple soft--gluon emission.
In hard scattering processes, the dynamics of the multiple
soft--gluon radiation is predicted by resummed pQCD 
\cite{resum}--\cite{resum4}. 

In this work, the Collins--Soper--Sterman (CSS) soft gluon
resummation formalism, developed for Drell--Yan pair (including $W$
and $Z$ boson) production \cite{css}, is extended to describe 
the production of photon pairs. 
This extension is similar to
the formalism developed for describing the
distribution of the leptons from vector boson decays \cite{wres}
because the final state
of the diphoton process
is also a color singlet state at LO.
Initial--state multiple soft--gluon emission in
the scattering subprocesses $q{\bar{q}},qg$ and $gg\to \gamma \gamma X$ is
resummed by treating the photon pair $\gamma \gamma $ similarly to the
Drell--Yan photon $\gamma ^{*}$.  In addition, there
are contributions 
in which one of the final photons is
produced through a long--distance fragmentation process.  An example
is $qg\to\gamma q$ followed by the fragmentation of
the final state quark $q\to \gamma X$.  
An earlier study of soft--gluon resummation effects in photon pair
production may be found in Ref.~\cite{qqres}.

The results of this overall calculation are compared with 
CDF \cite{cdfdata} and 
\D0 \cite{d0data} data
taken at the collider energy $\sqrt{S}=1.8$ TeV. A prediction for the
production rate and kinematic distributions of the diphoton pair
in proton--nucleon interactions at the fixed--target energy 
$\sqrt{S}=31.5$ GeV,
appropriate for the E706 experiment at Fermilab \cite{e706data},
is also presented.

Section 2 reviews some properties of the fixed order calculation of
the production rate and kinematics of
photon pairs.  Section 3 presents the soft gluon resummation formalism
and its application to diphoton production.  The numerical results
of this study, and a comparison with data, are presented in Sec. 4.
Finally, Sec. 5 contains a discussion of the results and conclusions.

\section{Diphoton Production at Leading and Next--to--Leading Order}

The leading order (LO) subprocesses
for diphoton production in hadron interactions are
of order $\alpha_{em}^2$, where $\alpha_{em}$ denotes the electromagnetic
coupling strength.  There are three classes of LO partonic contributions
to the reaction $h_1h_2\to \gamma \gamma X$, where $h_1$ and $h_2$
are hadrons, illustrated
in Fig.~(1a)--(1c).  The first (1a) is the short--distance $q\bar
q\to\gamma\gamma$ subprocess.  The second (1b) is the convolution of
the short--distance $qg\to\gamma q$ subprocess with the 
long--distance fragmentation of the final state quark $q\to \gamma X$.
This is a LO contribution since the hard scattering is of order
$\alpha_{em} \alpha_s$, while fragmentation is effectively of
order $\alpha_{em}/\alpha_s$.  Here, $\alpha_s$ denotes the QCD coupling
strength.
Class (1b) also includes the subprocess $q\bar q\to \gamma g$ convoluted
with the fragmentation $g\to \gamma X$.  
Finally, there are LO contributions (1c) involving subprocesses
like $qq\to qq$, where both final state quarks fragment $q\to\gamma X$.
The transverse momenta of the photons are denoted $\vec{p}_{T_1}$
and $\vec{p}_{T_2}$, and the transverse momentum of the pair is
$\vec{Q}_T = \vec{p}_{T_1} + \vec{p}_{T_2}$.  In the absence of
transverse momentum carried by the incident partons, the LO process
(1a) provides $\vec{Q}_T=0$.  With the added assumption of collinear
final--state fragmentation, (1b) provides 
$\vec{Q}_T = \vec{p}_{T_1} + \vec{p}_{T_2} = (1-z) \vec{p}_{T_1}$,
where photon 2 carries a fraction $z$ of the momentum of the final--state
quark. 
Given a lower limit on the magnitude of the
transverse momentum $p_T^\gamma$ of each photon, 
the total cross section at LO is finite.

The next--to--leading order (NLO) subprocesses for diphoton production
are of order $\alpha_{em}^2\alpha_s$.  One class of one--loop Feynman
diagrams (1d) 
contributes by interfering with the tree level diagram (1a).
Real gluon emission (1e) is also present at NLO.
The subprocess $qg\to\gamma\gamma q$ contains a singular piece (1f)
that renormalizes the lower order fragmentation (1b)
and a piece (1g) that is free of  final--state collinear singularities. 
Finally, subprocesses like $qq\to qq\gamma\gamma$ contain a
regular piece involving photon emission convoluted
with a fragmentation function (1h) and pieces that 
renormalize the double fragmentation process (1c).
The regular 3--body final state contributions from (1e), (1f),
and (1g)  provide
$\vec{Q}_T = -\vec{p}_{T_j}$, where $j$ represents the final--state
quark or gluon.
The full set of NLO contributions just described is free of final--state 
singularities, 
and the total integrated cross section at NLO is finite for
a finite lower limit on each $p_T^\gamma$.

Higher order calculations in $\alpha_s$ improve the
accuracy of predictions for total cross sections involving
quarks or gluons when only one hard
scale $Q$ is relevant. For $h_1h_2\to \gamma \gamma X$,
this scale can be chosen proportional to 
the invariant mass of the photon
pair, $Q=M_{\gamma \gamma }$, which is about equal to $2 p_T^\gamma$ 
for two well separated photons in the central rapidity region.
For kinematic distributions that
depend on more than one scale, a NLO calculation may be less
reliable.  One example is the distribution of the
transverse momentum of the photon pair, 
$Q_T=|\vec{Q}_{T}|$.
At fixed $Q$,
the behavior for small $Q_T$ of the NLO contribution to the differential 
cross section has
the form 
\begin{eqnarray}
\frac{d\sigma}{dQ_T^2}=\sigma_0\frac{\alpha_s}{\pi} 
\frac{1}{Q_T^2}\left[ a_1\ln \left( \frac{Q^2}{Q_T^2}\right)
+a_0 \right],
\label{Eq:NLO}
\end{eqnarray}
where $a_0$ and
$a_1$ are dimensionless constants, and $\sigma_0(Q)$ is the total LO cross
section calculated from the subprocess (1a). The structure
of Eq.~({\ref{Eq:NLO})
indicates that the fixed order QCD prediction is reliable when $Q_T\simeq Q$,
but becomes less reliable when $Q_T\ll Q$, where
$\ln(Q^2/Q_T^2)$ becomes large.  In the region $Q_T\ll Q$, the photon pair is
accompanied by soft and/or collinear gluon radiation. 
To calculate distributions like $\displaystyle\frac{d{\sigma}}{dQ_T^2}$
reliably in the region $Q_T \ll Q$, effects of multiple soft
gluon emission must be taken into account explicitly 
\cite{resum}--\cite{resum4}.
The contributions (1e) and (1g) 
exhibit singular behavior that can be tamed by resummation of
the effects of initial--state multiple soft--gluon radiation to all orders
in $\alpha_s$.
Other contributions that do not become singular as
$Q_T\to 0$ 
do not need to be
treated in this manner.
Fragmentation contributions like (1b) are found to be small in magnitude
after isolation restrictions are imposed on the energy of the hadronic
remnant from the fragmentation.  Therefore,
contributions like (1c) and (1h) are ignored in this work.  
Gluon fragmentation to a photon can be ignored, since its magnitude
is small.

The subprocess $gg\to\gamma\gamma$, involving a quark box diagram,
is of order $\alpha_{em}^2\alpha_s^2$.
While formally of even higher order than the NLO contributions 
considered so far,
this LO $gg$ contribution is enhanced by the size of the gluon 
parton distribution function.  
Consideration of the order $\alpha_{em}^2\alpha_s^3$ correction 
leads to resummation
of the $gg$ subprocess in a manner analogous to the $q\bar q$ resummation.

\section{The Soft Gluon Resummation Formalism}

To improve upon the prediction of Eq.~(\ref{Eq:NLO})
for the region $Q_T\ll Q$, perturbation theory can be applied using an
expansion parameter $\alpha_s^m\ln ^n(Q^2/Q_T^2)$, with $n=0,\ldots
,2m-1,$ instead of $\alpha_s^m$. 
The terms $\alpha_s^m\ln^n(Q^2/Q_T^2)$ represent the effects of soft
gluon emission at order $\alpha_s^m$.
Resummation of the singular part of the
perturbative series to all orders in $\alpha_s$ by Sudakov exponentiation
yields a regular differential cross section as $Q_T\to 0$. 

The differential cross section in the CSS resummation formalism for the
production of photon pairs in hadron collisions can be written as 
an integral in impact parameter $b$ space:
\begin{eqnarray}
&&{\frac{d\sigma (h_1h_2\to \gamma _1\gamma _2X)}{%
dQ^2\,dy\,dQ_T^2\,d\cos {\theta }\,d\phi }}={\frac 1{24\pi S}}\,{\frac
1{Q^2}}  \nonumber \\
&&~~\times \left\{ {\frac 1{(2\pi )^2}}\int d^2b\,e^{i{\vec{Q}_T}\cdot 
{\vec{b}}}\,\sum_{i,j}{\widetilde{W}_{ij}(b_{*},Q,x_1,x_2,\theta ,\phi
,C_1,C_2,C_3)}\,\widetilde{W}_{ij}^{NP}(b,Q,x_1,x_2)\right.  \nonumber \\
&&~~~~\left. +~Y(Q_T,Q,x_1,x_2,\theta ,\phi ,{C_4})\right\}.  
\label{master}
\end{eqnarray}
The variables $Q$, $y$, and $Q_T$ are the
invariant mass, rapidity, and transverse momentum of the photon pair in the
laboratory frame, while $\theta $ and $\phi $ are the polar and azimuthal
angle of one of the photons in the Collins--Soper frame \cite{CSFrame}.
The initial--state parton momentum fractions are defined as 
$x_1=e^yQ/\sqrt{S}$, and
$x_2=e^{-y}Q/\sqrt{S}$, and $\sqrt{S}$ is the center--of--mass (CM) 
energy of
the hadrons $h_1$ and $h_2$. 

The renormalization group invariant quantity $\widetilde{W}_{ij}(b)$ sums
the large logarithmic terms $\alpha_s^m\ln^{n}(b^2Q^2)$
to all orders in $\alpha_s$.
For a
hard scattering process initiated by the partons $i$ and $j$, 
\begin{eqnarray}
&&\widetilde{W}_{ij}(b,Q,x_1,x_2,\theta ,\phi ,C_1,C_2,C_3)= 
\exp \left\{ -{\cal S}_{ij}(b,Q{,C_1,C_2})\right\}  \nonumber \\
&&~\times \left[ {\cal C}_{i/h_1}(x_1){\cal C}_{j/h_2}(x_2)+ {\cal C}
_{j/h_1}(x_1){\cal C}_{i/h_2}(x_2)\right] 
{\cal F}_{ij}(\alpha_{em}(C_2 Q),\alpha_s(C_2 Q),\theta,\phi),
\label{Eq:WTwi}
\end{eqnarray}
where ${\cal F}_{ij}$ is a
kinematic factor that depends also on the coupling constants, and 
${\cal C}_{i/h}(x)$ denotes the convolution of the 
parton distribution function (PDF) $f_{a/h}$ 
(for parton $a$ inside hadron $h$) with 
the perturbative Wilson coefficient functions $C_{ij}^{(n)}$: 
\begin{equation}
{\cal C}_{i/h_1}(x_1)= \sum_a \int_{x_1}^1{\frac{d\xi _1}{\xi _1}}
\,C^{(n)}_{ia}\left( {\frac{x_1}{\xi _1}},b,\mu =\frac{C_3}b,C_1,C_2\right)
\;f_{a/h_1}\left( \xi _1,\mu =\frac{C_3}b\right) .
\end{equation}
The Sudakov exponent ${\cal S}_{ij}(b,Q{,C_1,C_2})$ in Eq.~(\ref{Eq:WTwi})
is defined as 
\begin{equation}
{\cal S}_{ij}(b,Q{,C_1,C_2})=\int_{C_1^2/b^2}^{C_2^2Q^2}{\frac{d{\bar{\mu}}^2
}{{\bar{\mu}}^2}}\left[ A_{ij}\left( \alpha_s({\bar{\mu}}),C_1\right) \ln
\left( {\frac{C_2^2Q^2}{{\bar{\mu}}^2}}\right) +B_{ij}\left( \alpha_s({\bar{
\mu}}),C_1,C_2\right) \right] .
\label{Eq:Sud}
\end{equation}
The coefficients
$A_{ij}$ and $B_{ij}$ and the functions $C_{ij}$
can be calculated perturbatively in powers of $\alpha_s/\pi $,
so that 
$\displaystyle A_{ij}=\sum_{n=1}^{\infty}(\alpha_s/\pi )^n A_{ij}^{(n)}$, 
$\displaystyle B_{ij}=\sum_{n=1}^{\infty}(\alpha_s/\pi )^n B_{ij}^{(n)}$,
and $\displaystyle C_{ij}=\sum_{n=0}^{\infty}(\alpha_s/\pi)^n C_{ij}^{(n)}$. 

The dimensionless constants $C_1,\;C_2$ and $C_3\equiv \mu b$ were
introduced in the solution of the renormalization group equations for 
$\widetilde{W}_{ij}$.  The constant $C_1$ determines the onset of 
nonperturbative
physics, $C_2$ specifies the scale of the hard scattering process, and $\mu
=C_3/b$ is the factorization scale at which the $C^{(n)}_{ij}$ functions are
evaluated. 
A conventional choice of the renormalization constants is 
$C_1=C_3=2e^{-\gamma _E}\equiv b_0$ and $C_2=C_4=1$ \cite{css},
where $\gamma_E$ is the Euler constant. 
These choices of the renormalization constants are used
in the numerical results of
this work because they eliminate large constant factors 
(depending on $C_1, C_2$ and $C_3$) in the Sudakov
exponent and in the $C_{ij}^{(n)}$ functions \cite{css}.

In Eq.~(\ref{master}), the impact parameter $b$ is to be integrated from 0
to $\infty $. However, for $b\ge b_{{\rm max}}$, which corresponds to an
energy scale less than $1/b_{{\rm max}}$, the QCD coupling $\alpha
_s(\bar\mu\sim 1/b)$ becomes so large that a perturbative calculation is no
longer reliable, and nonperturbative physics must set in. The nonperturbative
physics in this region is described by an empirically fit function 
$\widetilde{W}_{ij}^{NP}$ \cite{npfit,glenn}, 
and $\widetilde{W}_{ij}$ is evaluated
at a revised value of $b$,
$\displaystyle b_{*}={\frac b{\sqrt{1+(b/b_{{\rm max}})^2}}},$ where 
$b_{{\rm max}}$ is a phenomenological parameter used to separate 
long and short
distance physics. With this change of variable, $b_{*}$ never exceeds 
$b_{{\rm max}}$; $b_{\rm max}$ is a free parameter of 
the formalism \cite{css}
that can be constrained by other data (e.g. Drell--Yan).

The function $Y$ in Eq.~(\ref{master})
contains contributions in the full NLO
perturbative calculation that are
less singular than ${Q_T^{-2}}$ or
$Q_T^{-2}\ln({Q^2 /Q_T^2})$ as $Q_T \to 0$
(both the factorization and the
renormalization scales are chosen to be $C_4 Q$).
It is the difference between the exact
perturbative result to a given order and the result from
$\widetilde{W}_{ij}$ expanded to the same fixed order 
(called the asymptotic piece). 
The function $Y$ restores the
regular contribution in the fixed order perturbative calculation that is
not included in the resummed piece $\widetilde{W}_{ij}$. 
It does not contain 
a contribution from final--state fragmentation,
which is included separately as described in Sec. 3.2.

The CSS formula Eq.~(\ref{master}) contains many 
higher--order logarithmic terms, such that  
when $Q_T \sim Q$, the resummed differential cross section
can become negative in some regions of phase space.
In this calculation, the fixed--order prediction for the
differential cross section is used for $Q_T \grsim Q$
whenever it is larger than the prediction from Eq.~(\ref{master}).
The detailed properties of this matching prescription can be found in 
Ref. \cite{wres}.

\subsection{Resummation for the $q\bar{q}\to \gamma \gamma $ subprocess}

For the $q \bar q \to \gamma \gamma$ subprocess, the application of
the CSS resummation formalism is similar to the Drell--Yan case $q {\bar q}
^{~(^{\prime})} \to V^* \to \ell_1 {\bar \ell_2}$, where 
$\ell_1$ and $\ell_2$ are leptons produced through a 
gauge boson $V^{*}$ \cite{wres}. 
Since both processes are initiated by $q {\bar q}^{~(^{\prime})} $
color singlet states, the $A^{(1)}, A^{(2)}$ and $B^{(1)}$ functions in
the Sudakov form factor are identical to those of the
Drell--Yan case when each photon 
is in the central rapidity region with large
transverse momentum and is well
separated from the other photon.
This universality
can be understood as
follows. The invariants $\hat{s}$, $\hat{t}$ and $\hat{u}$ are defined
for the $q(p_1) \bar q(p_2) \to \gamma(p_3) \gamma(p_4)$ subprocess as 
\begin{equation}
\hat{s} = (p_1 + p_2)^2, ~~~~~~ \hat{t} = (p_1 - p_3)^2, ~~~~~~\hat{u} =
(p_2 - p_3)^2.
\end{equation}
The transverse momentum of each photon can be written as 
$p_T^\gamma=\sqrt{\hat{
t}\hat{u}/\hat{s}}$. When $p_T^\gamma$ is large, $\hat{t}$ and $\hat{u}$
must also be large, so the virtual--quark line connecting the two photons is
far off the mass shell, and the leading logarithms due to soft gluon
emission beyond the leading order can be generated only from the diagrams in
which soft gluons are connected to the incoming (anti--)quark.
To obtain the $B^{(2)}$ function, it is necessary to calculate  
beyond NLO, so it is not included in this calculation.  However, the
Sudakov form factor becomes more accurate when more terms are included
in $A_{ij}$ and $B_{ij}$.  
Since the universal functions $A^{(n)}_{ij}$ depend only on the flavor
of the incoming partons (quarks or gluons),
$A^{(2)}_{q\bar{q}}$ can be appropriated from Drell--Yan
studies, and its contribution {\it is} included in this paper.

To describe the effects of multiple soft--gluon emission, Eq.~(\ref{master})
can be applied, where $i$ and $j$ represent quark and
anti--quark flavors, respectively, and
${\cal F}_{ij} = \delta_{ij}(g_L^2+g_R^2)^2 (1 + \cos ^2\theta )/ (1 - \cos
^2\theta )$. The couplings $g_{L,R}$ are defined through the $q{\bar q}
\gamma$ vertex, written as 
$i\gamma _\mu \left[ g_L(1-\gamma _5)+g_R(1+\gamma _5)\right],$
with $g_L = g_R = e Q_f/2$, 
and $eQ_f$ is the electric charge of the incoming quark
($Q_{u} = 2/3, Q_{d} = -1/3$). The explicit forms of the $A$
and $B$ functions are:
\begin{eqnarray}
A^{(1)}_{q\bar q}(C_1) &=& C_F, \nonumber \\
A^{(2)}_{q\bar q}(C_1) &=& C_F \left[ \left( \frac{67}{36}-
\frac{\pi^2}{12}\right) N_C-
  \frac{5}{18}N_f -2\beta_1\ln\left(\frac{b_0}{C_1}\right) \right], 
\nonumber \\
B^{(1)}_{q\bar q}(C_1,C_2) &=& C_F \left[ -\frac{3}{2} -
2\ln\left(\frac{C_2 b_0}{C_1}
\right) \right], 
\end{eqnarray}
where $N_f$ is the number of light quark flavors,
$N_C=3$, $C_F=4/3$, and
$\beta_1=(11N_C-2N_f)/12$.

To obtain the value of the total cross section to NLO,
it is necessary to include the Wilson coefficients
$C_{ij}^{(0)}$ and
$C_{ij}^{(1)}$.  
These can be derived
from the full set of LO contributions and
NLO corrections to $\gamma \gamma$ production \cite{nlo}.
After the leading order and the one--loop virtual corrections
to $q\bar q\to\gamma\gamma$ and the tree level contribution from 
$q\bar q \to\gamma\gamma g$ are included, the coefficients are:
\begin{eqnarray}
C_{jk}^{(0)}(z,b,\mu ,{\frac{C_1}{C_2}}) &=&\delta _{jk}\delta ({1-z}), 
\nonumber \\
C_{jG}^{(0)}(z,b,\mu ,{\frac{C_1}{C_2}}) &=&0, \nonumber \\
C_{jk}^{(1)}(z,b,\mu ,{\frac{C_1}{C_2}}) &=&\delta _{jk}C_F
\left\{ \frac{1}{2}(1-z)-\frac{1}{C_F}\ln\left(\frac{\mu b}{b_0}\right) 
P_{j\leftarrow k}^{(1)}(z)\right. \nonumber \\
&&\left. +\delta (1-z)\left[ -\ln ^2\left( {\frac{C_1}{{b_0C_2}}}
e^{-3/4}\right) +{\frac{{\cal V}}4}+{\frac 9{16}}\right] \right\}.
\label{eq:quarks}
\end{eqnarray}
After factorization of the final--state collinear
singularity, as described below,
the real emission subprocess
$qg\to \gamma\gamma q$ yields:
\begin{eqnarray}
C_{jG}^{(1)}(z,b,\mu ,{\frac{C_1}{C_2}}) &=&{\frac 12}z(1-z)-\ln \left(
\frac{\mu b}{b_0} \right) P_{j\leftarrow G}^{(1)}(z).
\label{eq:gluons}
\end{eqnarray}
In the above expressions, the splitting kernels \cite{dglap}
are 
\begin{eqnarray}
P_{{j\leftarrow k}}^{(1)}(z) &=&C_F\left( {\frac{1+z^2}{{1-z}}}\right)
~~~{\rm and}  \nonumber \\
P_{{j\leftarrow G}}^{(1)}(z) &=&{\frac 12}\left[ z^2+(1-z)^2\right].
\label{eq:DGLAP}
\end{eqnarray}
For photon pair production, the function
${\cal V}$ is
\begin{eqnarray}
\label{eq:Vgg}
{\cal V}_{\gamma \gamma }&=&-4+{\frac{\pi ^2}3}+
{\frac{\hat{u}\hat{t}}{\hat{u}
^2+\hat{t}^2}}\left( F^{virt}(v)-2\right), \nonumber \\
F^{virt}(v) &=&\left( 2+{\frac v{1-v}}\right) \ln ^2(v)+\left( 2+{\frac{1-v}v
}\right) \ln ^2(1-v)  \nonumber \\
&&+\left( {\frac v{1-v}}+{\frac{1-v}v}\right) \left( \ln ^2(v)+\ln ^2(1-v)-3+
{\frac{2\pi ^2}3}\right)  \nonumber \\
&&+2\left( \ln (v)+\ln (1-v)+1\right) +3\left( {\frac v{1-v}}\ln (1-v)+{
\frac{1-v}v}\ln (v)\right), \nonumber \\
\end{eqnarray}
where $v=-\hat{u}/\hat{s}$, and $\hat{u}=-\hat{s}(1+\cos \theta)/2$ in the 
$q \bar{q}$ center--of--mass frame.
Because of Bose symmetry, $F^{virt}(v)=F^{virt}(1-v)$. 
A major difference
from the Drell--Yan case (${\cal V}_{DY}=-8+\pi ^2$)
is that ${\cal V}_{\gamma \gamma }$ depends on the
kinematic correlation between the initial and final states
through its dependence on $\hat{u}$ and $\hat{t}$.

The nonperturbative function used in this study is the empirical fit
\cite{glenn}
\begin{eqnarray}
\widetilde{W}_{q\overline{q}}^{NP}(b,Q,Q_0,x_1,x_2) = {\rm exp}\left[
-g_1b^2-g_2b^2\ln \left( {\frac Q{2Q_0}}\right) -g_1g_3b\ln {(100x_1x_2)}
\right],
\label{Eq:WNP}
\end{eqnarray}
where $g_1=0.11_{-0.03}^{+0.04}~{\rm GeV}^2$, $g_2=0.58_{-0.2}^{+0.1}~{\rm 
GeV}^2$, $g_3=-1.5_{-0.1}^{+0.1}~{\rm GeV}^{-1}$, and $Q_0=1.6~{\rm GeV}$.
(The value $b_{max}=0.5~{\rm GeV}^{-1}$ was used in determining the above
$g_i$'s and for the numerical results
presented in this paper.) These values were fit for the 
{\tt CTEQ2M} parton distribution function,
with the conventional choice of the renormalization constants, i.e.
$C_1=C_3=b_0$ and $C_2=1$.
In principle, these coefficients should be refit
for the {\tt CTEQ4M} distributions \cite{cteq} used in this study.
The parameters of Eq.~(\ref{Eq:WNP}) were determined from Drell--Yan
data.   It is assumed that the same values should be applicable
for the $\gamma\gamma$ final state.

\subsection{Contributions From $qg$ Subprocesses}

As described in Sec. 2, the complete NLO calculation of
diphoton production in hadron collisions includes photons from
long--distance fragmentation processes like (1b) and
short--distance processes like (1f) and (1g).
The latter processes yield a regular 3--body final state contribution, 
while the former
describes a photon recoiling against a collinear quark and photon.

The singular part of the 
squared amplitude of the $q(p_1)g(p_2)\to \gamma (p_3)\gamma
(p_4)q(p_5)$ subprocess can be factored into
a product of the squared amplitude
of $q(p_1)g(p_2)\to \gamma (p_3)q(p_{4+5})$ and the splitting kernel for
$q(p_{4+5})\to \gamma (p_4)q(p_5)$.
In the limit that
the emitted photon $\gamma (p_4)$
is collinear with the final state quark $q(p_5)$:
\begin{eqnarray}
\lim_{p_4\parallel p_5} && \left| 
{\cal M}\left( q(p_1)g(p_2)\to \gamma
(p_3)\gamma (p_4)q(p_5)\right) \right| ^2 = \nonumber \\
&&{\frac{e^2}{p_4 \cdot p_5}}P_{\gamma\leftarrow q}^{(1)}\left( z\right)
\left| {\cal M}\left( q(p_1)g(p_2)\to \gamma (p_3)q(p_{4+5})\right)
\right| ^2.
\label{subtraction}
\end{eqnarray}
A similar result holds 
when $p_3$ and $p_5$ become collinear and/or the quark is replaced with
an anti--quark.
Conventionally, the splitting variable $z$ is the light--cone
momentum fraction of the emitted photon with respect to the
fragmenting quark, $z=p_4^{+}/(p_4^{+}+p_5^{+})$,
where $p_i^{+}=\left( p_i^{(0)}+p_i^{(3)}\right) /
\sqrt{2}$. 
($p_i^{(0)}$ is the energy and $p_i^{(3)}$ is the longitudinal 
momentum component along the moving direction of the 
fragmenting quark in the $qg$ center-of-mass frame.) 
Alternatively, since
the final state under consideration contains only a fragmenting
quark and a spectator,
a Lorentz invariant splitting variable can be defined as:
\cite{Catani-Seymour}
\begin{equation}
\tilde{z}=1-\frac{p_i\cdot p_k}{p_j\cdot p_k+p_i\cdot p_k+p_i\cdot p_j}.
\end{equation}
In this notation, $i=5$ is the fragmentation quark, $j$ is the fragmentation
photon, and $k$ is the prompt spectator photon.
When the pair $ij$ becomes collinear, $\tilde z$ becomes  
the same as the light--cone momentum fraction $z$ carried by the photon.  
Aside from
the color factor, $P_{\gamma\leftarrow q}^{(1)}(z)$ in 
Eq.~(\ref{subtraction})
is the usual DGLAP splitting kernel for $q\to gq$ 
\begin{eqnarray}
P_{\gamma \leftarrow q}^{(1)}(z) &=&\left( \frac{1+(1-z)^2}{z}\right).
\end{eqnarray}

The regular contribution $qg\to\gamma\gamma q$ (1f) is defined by
removing the final--state, collinear singularity from the full
amplitude of the partonic subprocess.  
The matrix element squared for (1f) 
can be written \cite{Catani-Seymour}:
\begin{eqnarray}
\left| {\cal M}\left( qg\to\gamma\gamma q \right) \right|^2_{reg} &=& 
\nonumber \\
\left| {\cal M}\left( qg\to\gamma\gamma q \right) \right|^2_{full} &-& 
{{{e^2}\over{p_4 \cdot p_5}}}P_{\gamma\leftarrow q}^{(1)}
\left(\tilde z\right)
\left| {\cal M}\left( q(p_1)g(p_2)\to \gamma (p_3)q(p_{4+5})\right)
\right| ^2.
\label{regular}
\end{eqnarray}
After the final--state collinear singularity is subtracted,
the remainder expresses
the regular 3--body final state contribution $\gamma\gamma q$.
This remainder, as shown in (1g), contains
terms that diverge when $Q_T\to 0$ which should be regulated
by renormalizing the parton distribution at the NLO.  The contribution
from this divergent part is included in the resummed $q\bar q$
cross section in $C^{(1)}_{jG}$, as shown in Eq.~(\ref{eq:gluons}).
The part
that is finite as $Q_T\to 0$ is included in the function $Y$.
When $Q_T \grsim Q$,
Eq.~(\ref{regular}) describes the NLO contribution from the 
$qg\to\gamma\gamma q$
subprocess to the $Q_T$ distribution of the photon pair.
The subtracted final--state collinear singularity from the NLO
$qg\to\gamma\gamma q$ subprocess is absorbed into the fragmentation
process (1b).

\subsection{Fragmentation Contributions}

Final--state photon fragmentation functions $D_{\gamma/i}(z,\mu_F^2)$ are
introduced in an analogous manner to initial--state parton distribution
functions $f_{i/h_1}(x,\mu_I^2)$.  Here, $z (x)$ is the light--cone momentum
fraction of the fragmenting quark (incident hadron) carried by the 
photon (initial--state parton), and $\mu_F (\mu_I)$ is
the final state (initial state) fragmentation (factorization) scale.
The parton--level cross section for
the fragmentation contribution (1b) is
evaluated from the general expression for a hard scattering
to a parton $m$, which then fragments to a photon:
\begin{equation}
d\hat{\sigma} = \frac{1}{2\hat{s}} |{\cal M}(p_1p_2\to p_3\ldots p_m) |^2
d^{(m)}[PS] dz D_{\gamma/m}(z,\mu_F^2).
\end{equation}
Here, $\cal M$ is the matrix element for the hard scattering subprocess,
$d^{(m)}[PS]$ is the $m$--body phase space, 
and an integral is performed over the photon
momentum fraction $z$ weighted by the 
fragmentation function $D_{\gamma/m}(z,\mu_F^2)$.
Since fragmentation is computed here to LO only, the
infrared divergences discussed by Berger, Guo and Qiu are
not an issue \cite{BGQiu}.

The fragmentation function $D_{\gamma\leftarrow q}$ obeys an 
evolution equation, and
the leading--logarithm, asymptotic solution 
$D^{LL}_{\gamma/q}$ is \cite{nlo}: 
\begin{eqnarray}
\label{dfunction}
D^{LL}_{\gamma/q}(z,\mu_F^2) & = &
{\alpha_{em} \over 2\pi}\ln\left({\mu_F^2 \over \Lambda_{QCD}^2}\right) 
D^{(1)}_{\gamma\leftarrow q}(z),\nonumber \\
zD_{\gamma\leftarrow q}^{(1)}(z) & = & 
\frac{Q_q^2(2.21-1.28z+1.29z^2)z^{0.049}}{1-1.63\ln (1-z)} +
0.0020(1-z)^{2.0}z^{-1.54}, 
\end{eqnarray}
where $\Lambda_{QCD}$ is the QCD scale for four light quark flavors. 
As shown in Fig. \ref{Fig:Frag},
the collinear approximation made in defining $D_{\gamma\leftarrow q}$ 
leads to
kinematic distributions with an unrealistic sensitivity to
kinematic cuts, such as cuts to define an isolated photon.

The Monte Carlo showering method goes beyond the collinear approximation
used in solving the evolution equation for the fragmentation function 
$D_{\gamma\leftarrow q}$.  
In Monte Carlo calculations,
the probability for photon
emission is determined
from the splitting function $P_{\gamma\leftarrow q}(z)$, which is a
collinear approximation.  However, the kinematics are treated
by assigning a virtuality to the fragmenting quark whose
value lies between the
hard scale of the process and a phenomenological cutoff $\sim 1$ GeV.
This cutoff replaces the parameter $\Lambda_{QCD}$ in Eq.~(\ref{dfunction}).
Most importantly, gluon emission can be incorporated into the
description of final state fragmentation.  Because there is no
collinear approximation in the kinematics, kinematic distributions
do not exhibit the unrealistic behavior of the parton--level calculation.
The ``correctness'' of either approach can be judged only after a
careful comparison of their respective predictions.

The collinear approximation becomes an issue because of the experimental
definition of isolated photons.
Experimentally, an isolation cut is necessary to
separate prompt photons from various hadronic backgrounds,
including $\pi^\circ$ and $\eta$ meson decays.
The separation between a particle $j$ and the photon is expressed as
$R_j = \sqrt{(\eta-\eta_j)^2 + (\phi-\phi_j)^2}$,
where the coordinates $\eta (\eta_j)$ and $\phi (\phi_j)$ 
are the pseudorapidity
and azimuthal angle of the photon (particle $j$).
At hadron colliders, the standard isolation
criterion is that the sum of excess transverse energy $E_T$ 
contained inside a
cone of size $R_0$ centered on the photon
candidate is below a cutoff $E_T^{iso}$, $\displaystyle 
\sum_{R_j<R_0} E_T^j <
E_T^{iso}$.  
The sum is over each
particle $j$.
Since the resummed CSS piece of the final state cross section
describes the radiation of multiple soft gluons
approximately collinear with the incident partons,
it produces only isolated photons.  
For NLO $\gamma\gamma j$ final states (1e), (1f), and
(1g), where there is only one extra parton $j=q$ or $g$, 
isolation enforces a separation $R_{j} \ge R_0$, provided that
$p_{T_j} > E_T^{iso}$. 
Above $Q_T=E_T^{iso}$, the perturbative corrections
contained in the function $Y$ are affected by isolation.
On the other hand,
because of the collinear approximation, the parton--level
fragmentation calculation based on
Eq.~(\ref{dfunction}) does not depend on the isolation cone $R_0$;
the hadronic remnant of the fragmentation (1b) {\it always} 
satisfies $R<R_0$.
Hence, for this case, $\vec{Q}_T = (1-z) \vec{p}_{T_1}$, and the isolation
cut reduces to a step function requirement $\theta(E_T^{iso}-Q_T)$.

The parton--level calculation of the fragmentation
contribution at the Tevatron 
based on the fragmentation function $D_{\gamma\leftarrow q}(z,\mu_F^2)$
has been compared with a Monte Carlo estimate
based on {\tt PYTHIA} \cite{pythia}.  
For the 
parton--level calculation,
the scale $\mu_F=M_{\gamma\gamma}$ is used.  
For the {\tt PYTHIA} calculation,
the scale is $\mu_F=\sqrt{\hat{s}}$, and 
hadronization is not performed, so that no photons arise from 
$\pi^\circ$ or $\eta$ meson decays, for example.  For this comparison,
the invariant mass $\sqrt{\hat{s}}$ of 
the hard--scattering subprocess is limited
to $20<\sqrt{\hat{s}}<50$ GeV in both approaches, 
and the photons are required to satisfy
$p_T^\gamma>5$ GeV and $|\eta^\gamma|<2$.
These kinematic cuts are chosen to increase the
statistics of the {\tt PYTHIA} calculation, while reflecting the
kinematic region of interest for a comparison with data.  
{\tt PYTHIA} can simulate the QED and QCD showering of the final--state quark
as well as the QCD showering of the initial--state quark and gluon.
To isolate the effect of initial--state gluon
radiation, {\tt PYTHIA} calculations
were performed with and without the QCD initial--state radiation (i.e. by
preventing space--like showering).
In neither case is initial--state QED radiation simulated.  It
is possible for the partons produced in initial--state showering
to develop time--like showering.  Any photons produced from this
mechanism are discarded, since they are formally of higher--order
than the contributions considered here.  Such contributions, however,
might be necessary to understand photon pairs with small invariant
mass and small $Q_T$.

Figure~\ref{Fig:Frag} is a comparison of kinematic quantities
from the parton--level and Monte Carlo calculations.  The left--side
of Fig.~\ref{Fig:Frag}
shows the $Q_T$--distribution for the parton--level (solid),
{\tt PYTHIA} with initial--state radiation of gluons (short--dashed), 
and  {\tt PYTHIA} without initial--state radiation (long--dashed) 
calculations.  Each curve is plotted twice, with and without
an isolation cut $E_T^{iso}=4$ GeV and $R_0=0.7$.  Before the
isolation cut, the total parton--level fragmentation
cross section is approximately
50\% higher than the Monte Carlo cross section.  After isolation,
the total cross sections are in good agreement, even though the
parton--level calculation is discontinuous at $Q_T=E_T^{iso}$.
The effect of initial--state gluon radiation in the
{\tt PYTHIA} calculation is to increase 
$Q_T$ without compromising the isolation of the photons.

The right--side of Fig.~\ref{Fig:Frag} 
shows the distribution of the light--cone
momentum fraction $z$ of the quark carried by the fragmentation photon
(for this figure, $z$ is defined in the laboratory frame).
After isolation, the parton--level contribution is limited 
to $z>0.55$ by kinematics, whereas the Monte Carlo contribution
is more uniformly distributed between 0 and 1.  For the {\tt PYTHIA}
result, $z$ is calculated with respect to the final state quark
{\it before} showering.  In the showering process, some energy--momentum
can be exchanged between the final state prompt photon and the
fragmenting quark, since the quark is assigned a virtuality.
As a result, the effective $z$ value can extend
beyond the naive limit $z=1$.

The conclusions of this comparison are: (1) after isolation, 
the total cross sections from
the parton--level and Monte Carlo fragmentation calculations
are in good agreement, and (2) the Monte Carlo
kinematic distributions (e.g. $Q_T$ and $z$)
are not very sensitive to the isolation cut.
For these reasons, the Monte Carlo estimate with 
initial--state radiation is used to account for the (1b) 
contribution in the final results.  
Furthermore, with initial--state radiation,
the {\tt PYTHIA} calculation
includes the leading effects of
a full resummation calculation of the $qg\to\gamma q$ process.  
It is approximately equivalent
to performing a resummation calculation in the CSS formalism
with quantities $A^{(1)}$ and $B^{(1)}$ calculated for a $qg$ 
initial state and the LO Wilson function.

One final comparison was made with the Monte Carlo calculation
by treating the subtracted term in Eq. (16),
with $P^{(1)}$ replaced by $D^{(1)}$ defined in Eq. (18),
as a 3--body matrix element. The collinear divergence was regulated by
requiring a separation $R_0$ between the photon and
quark remnant for all $Q_T$.  This calculation agrees with
{\tt PYTHIA} in the shape and normalization of various distributions,
except when $Q_T < E_T^{iso}$, where there is a substantial difference.

\subsection{Resummation for the $gg\to \gamma \gamma $ subprocess}

A resummation calculation for the $gg\to \gamma \gamma $
subprocess is included in the theoretical prediction. The 
LO contribution comes from one--loop box diagrams of
order $\alpha _{em}^2\alpha_s^2$ in perturbative QCD. 
At present, a full NLO calculation,
of ${\cal O}(\alpha_{em}^2\alpha_s^3)$,
 for this process is not available.
Nevertheless, the resummation technique can be applied to resum part of the
higher order contributions and improve the theoretical prediction. 
The exact NLO
$gg\to \gamma \gamma g$ calculation must include gluon emission from the
internal quark lines of the box diagram, thus generating pentagon diagrams.
However, such diagrams do not generate large logarithms when the final state
photons have large transverse momentum, are in the central rapidity
region, and are well separated from each other.
All the large logarithms originate from the
diagrams with soft gluons coupling to the initial--state  
gluons. 
Similarly, the exact NLO $qg\to\gamma\gamma q$ calculation,
of ${\cal O}(\alpha_{em}^2\alpha_s^3)$,
must include contributions
involving a box diagram with one incoming gluon off shell.
Large logarithms only arise from soft gluon emission off the
initial--state quark or gluon.
The leading logarithms due to initial--state radiation are universal,
and the $A^{(1)}$
function calculated for the resummed $gg\to H$ 
process \cite{higgsres,cpyhiggs} 
or the color singlet part of the $gg\to Q\bar Q$ process \cite{resum4}
can be applied directly to the resummed $gg\to\gamma\gamma$ calculation, 
since these subprocesses have
the same QCD\ color structure.

When the transverse momentum of the photon pair is much smaller than its
invariant mass, i.e. $Q_T\ll Q$, and each photon
has large transverse momentum, then the box diagram of the hard
scattering subprocess $gg\to \gamma \gamma $ can be approximated as a
point--like interaction (multiplied by a form factor which
depends on $\hat{s},\hat{t}$ and $\hat{u}$).
This approximation ignores pentagon diagrams in the
$gg\to\gamma\gamma g$ subprocess and the virtuality of intermediate
quarks in the $qg\to \gamma\gamma q$ subprocess.
It does not have the complete structure of the
hard process, but it does contain the most important logarithmic
terms from initial state gluon radiation.
Under such an approximation, the subleading logarithmic terms 
associated with $B^{(1)}$, 
$A^{(2)}$, and $C^{(1)}$ of Eqs.~(4) and (5)
can be included in the resummation calculation.
These functions were calculated for the $gg\to H$ process 
\cite{higgsres,cpyhiggs}. 
Without a complete ${\cal O}(\alpha_{em}^2\alpha_s^3)$
calculation, the exact Wilson coefficient
function $C^{(1)}$ is not known.
Since part of the exact $C^{(1)}$ function must include the
piece for the $gg\to H$ process, it is included to estimate
the possible NLO enhancement to the production rate
of the $gg$ subprocess.
After these ingredients are incorporated
into Eq. (\ref{master}),
the resummed kinematics of the photon pair from the
$gg\to \gamma \gamma $ subprocess can be obtained.
The distribution of the individual
photons can be calculated 
approximately from the LO\ angular dependence of the box
diagram.

The above approximation certainly fails when $Q_T$ is of the order of $Q$.
In the absence of a complete ${\cal O}(\alpha_{em}^2\alpha_s^3)$
 calculation of the $gg\to \gamma
\gamma g$ and $qg\to\gamma\gamma q$ subprocesses, 
it is not possible to estimate the uncertainties
introduced by the approximation. In the limit of $Q_T\ll Q$, the
approximation
should be reliable,
since the soft gluon approximation is applicable. 
In the same spirit, the approximate function $Y$
for photon pair production is taken from the results of the 
perturbative piece for
the $gg\to Hg$ and $g\qgen\to H\qgen$
processes \cite{higgsres,cpyhiggs}.

\smallskip In summary, the resummed distributions of the
photon pair from the $gg$ subprocess in the region of $Q_T\ll Q$ can
be described
by Eq. (\ref{master}),
with $i=j=g$, and
  ${\cal F}_{gg} = N_C \left| {\cal M}_{gg \to \gamma\gamma}(s,t,u) 
  \right|^2/2^{12}$.  Here,
$\left| {\cal M}_{gg\to\gamma\gamma}(s,t,u) \right|^2$ is the 
  absolute square of the invariant amplitude 
of the $gg \to \gamma\gamma$ subprocess 
\cite{box}
  summed over spins, colors, and the fermion flavors in the box loop, but
  without the initial--state color $(1/8^2)$, spin $(1/2^2)$ average, and
  the final--state identical particle $(1/2)$ factors.
The $A$ and $B$ functions
used in the calculation for the $gg$ initial state are 
\begin{eqnarray}
A^{(1)}_{gg}(C_1) &=& C_A=3,  \nonumber \\
A^{(2)}_{gg}(C_1) &=& \frac{C_A}{C_F} A^{(2)}_{q\bar q}(C_1),  \nonumber \\
B^{(1)}_{gg}(C_1,C_2) &=&2\left[ 3\ln \left( \frac{C_1}{C_2b_0}
\right) -\beta _1\right].
\end{eqnarray}
The LO and NLO Wilson coefficients, extracted from the $gg\to H$
subprocess, are:
\begin{eqnarray}
C_{gg}^{(0)}\left( z,b;\frac{C_1}{C_2};\mu \right) &=&\delta (1-z), 
\nonumber \\
C_{qg}^{(0)}\left( z,b;\frac{C_1}{C_2};\mu \right)  &=& 0 \nonumber \\
C_{gg}^{(1)}\left( z,b;\frac{C_1}{C_2};\mu \right) &=&-\ln \!\left( 
\frac{\mu b}{b_0}\right) P_{g\leftarrow g}(z)  
+\delta (1-z)\left\{ \frac{11}{4}+\frac{3\pi^2}{4} \right. \nonumber \\
& & \left. -3\ln^2\left(\frac{C_1}{C_2b_0}\right) 
+3\ln\left( \frac{C_1}{C_2b_0}\right) 
+(2\beta_1-3)\ln \left( \frac{\mu b}{b_0}\right) \right\}, \nonumber \\
C_{qg}^{(1)}\left( z,b;{\frac{C_1}{C_2}};\mu \right)  &=& 
-\ln\left(\frac{\mu b}{b_0}\right)P_{g\leftarrow q}(z) + \frac{2}{3}z.
\end{eqnarray}

Since the NLO pentagon and off--shell box diagram calculations are not 
included, the Wilson
coefficients $C_{ij}^{(1)}$ are expected to predict accurately the total
cross section only when $Q_T\ll Q$, the transverse momenta of the 
individual photons are
large, and their rapidities are small. 
Under the {\it approximation} made above, 
the resummed $gg$ result increases the
integrated rate by about a factor of 2, for kinematic cuts typical 
of the Tevatron,
as compared to the lowest order (one--loop calculation) perturbative result.
This comparison suggests that the full NLO
contribution of the $gg$ initiated subprocess is large. Because it is
necessary to impose the condition $Q_T<Q$ to make the above approximations
valid, the $gg$ resummed result presented in this work probably underestimates
the rate when $Q_T$ is large or the separation of the azimuthal angle 
($\Delta\phi$) between the two photons is small. This deficiency
can be improved only
by a complete ${\cal O}(\alpha_{em}^2\alpha_s^3)$ calculation.

At the Tevatron, the $gg$ contribution is important when the invariant
mass ($M_{\gamma \gamma }=Q$) of the two photon pair is small. Because of
the approximation made in the $gg$ calculation beyond the LO, 
the prediction will be more reliable for the data with larger $Q$. A more
detailed discussion is presented in the next section.

The full calculation of the $gg$ contribution in the CSS formalism 
depends also
upon the choice of nonperturbative functions. However, the best fits to
the parametrizations are performed for $q\bar{q}$ processes \cite{npfit,glenn}. 
Two assumptions were studied: $(i)$ the nonperturbative functions
are
truly universal for $q\bar q$ and $gg$ processes, and
$(ii)$ the nonperturbative functions obey the same renormalization
group properties as the $A$ functions for each type of process (which are
universal for all $q\bar{q}$ or $gg$ subprocesses), so the coefficient
of the $\ln \left( \frac Q{2Q_0}\right) $ term in the nonperturbative
function Eq.~(\ref{Eq:WNP}) 
is scaled by $C_A/C_F$ relative to that of the $q\bar q$ process.
Specifically, the different assumptions are:
\begin{eqnarray}
(i) \widetilde{W}_{gg}^{NP}(b,Q,Q_0,x_1,x_2) &=& \widetilde{W}^{NP}_{q\bar q}
(b,Q,Q_0,x_1,x_2)
\nonumber \\
(ii) \widetilde{W}_{gg}^{NP}(b,Q,Q_0,x_1,x_2) &=&
\widetilde{W}_{q\overline{q}}^{NP}(b,Q,Q_0,x_1,x_2) 
(g_2 \to {C_A\over C_F} g_2). \nonumber \\
\label{Eq:ggnp}
\end{eqnarray}
The numerical values of $g_1, g_2,$ and $g_3$ are listed following
Eq.~(\ref{Eq:WNP}).  These two assumptions do not exhaust all 
possibilities,
but ought to be representative of reasonable choices.
Choice $(ii)$ is used for the results presented in this paper.
The effect of different choices is discussed in Sec.~4.

\section{Numerical Results}

\subsection{Tevatron Collider Energies}

Two experimental collaborations at the Tevatron $p\bar p$ collider
have collected diphoton
data at $\sqrt{S}=1.8$ TeV:  CDF \cite{cdfdata}, with 84 pb$^{-1}$, 
and \D0~\cite{d0data}, with 81 pb$^{-1}$.  
The kinematic cuts
applied to the resummed prediction for comparison with the CDF data 
are  $p_T^\gamma > 12$ GeV and $|\eta^{\gamma}|<0.9$.  For \D0, the
kinematic cuts are $p_T^{\gamma_1} > 14$ GeV  and 
$p_T^{\gamma_2} > 13$ GeV, 
and $|\eta^{\gamma}|<1$.
For CDF, an isolation cut for each photon of
$R_0=0.7$ and $E_T^{iso}=4$ GeV is applied;
for \D0, the cut is $R_0=0.4$ and $E_T^{iso}=2$ GeV.

Other ingredients of the calculation are: $(i)$ the {\tt CTEQ4M}
parton distribution functions, 
$(ii)$ the NLO expression for $\alpha_s$,
$(iii)$ the NLO expression for $\alpha_{em}$, and 
$(iv)$ the nonperturbative
coefficients of Ladinsky--Yuan \cite{glenn}. 

The predictions for the CDF cuts and a comparison to the data are
shown in Figs.~\ref{Fig:CDF0}--\ref{Fig:CDF2}.
Figure~\ref{Fig:CDF0} shows the distribution of the photon pair
invariant mass,
$d\sigma/dM_{\gamma\gamma}$ vs. $M_{\gamma\gamma}$.
  The dashed--dot curve represents the
resummation of the $gg$ subprocess, which is the
largest contribution for $M_{\gamma\gamma}\lesssim  30$ GeV.
The long--dashed curve represents the full $q\bar q$
resummation, while the short--dashed curve is a similar calculation
with the gluon parton distribution function artificially set to zero.  
Schematically, there are contributions
to the resummed calculation
that behave like $q\to g q_1\otimes q_1\bar q\to\gamma\gamma$
and $g\to \bar q q_1 \otimes q_1\bar q\to\gamma\gamma$.  
These contributions are contained  in the terms proportional
to $P_{j\leftarrow k}^{(1)}(z)$ in Eq.~(\ref{eq:quarks}) and
   $P_{j\leftarrow G}^{(1)}(z)$ in Eq.~(\ref{eq:gluons}),
respectively.  
The full $q\bar q$ resummation contains both the $q\bar q$ and $qg$
contributions.
The short--dashed curve is calculated by setting
$C_{jG}^{(1)}=0$ and retaining only the $q\bar q$ contribution in
the function $Y$.
Since
the short--dashed curve almost saturates the full 
$q\bar q+qg$ contribution,
except at large $Q_T$ or small $\Delta\phi$, the
$qg$ initiated subprocess is not important at the Tevatron in most
of phase space for the cuts used.
The fragmentation contribution is denoted by the dotted line.
The sum of all contributions including fragmentation
is denoted by the solid line.
After isolation,
the fragmentation contribution is
much smaller than ``direct'' ones, but
contributes $\simeq 10\%$ near the peak.  
The uncertainty in the contribution
of the fragmentation process can be estimated by
comparing the Monte Carlo result with a parton--level calculation,
as shown in Fig.~\ref{Fig:Frag}. 

Figure \ref{Fig:CDF1} shows the distribution of the
transverse momentum of the photon pair,
$d\sigma/dQ_T$ vs. $Q_T$.
Over the interval $5\lesssim Q_T \lesssim 25$ GeV, the contribution
from the $gg$ subprocess is comparable to the $q\bar q+qg$ subprocess.
The change in slope near $Q_T=20$ GeV arises from the 
$gg$ subprocess (dot--dashed line) for which
$Q_T \stackrel{\scriptscriptstyle<}{\scriptscriptstyle\sim}
M_{\gamma\gamma}$ is required in our approximate calculation.
The peak near $Q_T\simeq 1.5$ GeV is provided mostly by the
$q\bar q+qg$ (long--dashed line) subprocess.
In general, the height and breadth of the peak in the $Q_T$
distribution depends on the details of the nonperturbative function
in Eq.~(\ref{master}).
The effect of different nonperturbative
contributions may be estimated if the parameter
$g_2$ is varied by $\pm 2\sigma$. For $Q_T > 10$ GeV,
the distribution is not sensitive to this variation. 
The height and the width (half--maximum) of the peak 
change by approximately 20\% and 35\%, respectively, but
the integrated rate from 0 to 10 GeV is almost constant.
The peak of the distribution (which is
below 5 GeV), shifts approximately $+0.5$ GeV or $-0.6$ GeV for a
$+2\sigma$ or $-2\sigma$ variation. The mean $Q_T$ for $Q_T<10$ GeV
shifts at most by $0.4$ GeV.  
For $gg$ resummation, it is not clear which parametrization of
the nonperturbative physics should be used.
However, the final effect of the two different parametrizations
outlined in Eq.~(\ref{Eq:ggnp}) is minimal,
shifting the mean $Q_T$ for $Q_T<40$ GeV by about 2.0 GeV.
The parametrization $(ii)$ is used in the final results, so that
the coefficient $g_2$ is scaled by $C_A/C_F$ relative to the
$q\bar q$ nonperturbative function.

Figure~\ref{Fig:CDF2} shows 
$d\sigma/d\Delta\phi$ vs. $\Delta\phi$, where $\Delta\phi$ is the azimuthal
opening angle between the two photons. The change in slope near 
$\Delta\phi=\pi/2$ is another manifestation of the approximations made in the 
treatment of the
$gg$ contribution (dot--dashed line).
The height of the distribution near $\Delta\phi\simeq\pi$ is also
sensitive to the details of the nonperturbative function.

In the absence of resummation or NLO effects, the $gg$ box contribution
supplies $\vec{Q}_T=0$ and $\Delta\phi=\pi$.
In this calculation, as explained earlier,
the NLO contribution
for the $gg$ subprocess is handled in an approximate fashion.
For the cuts listed above,
the total cross section from the complete $gg$ resummed calculation,
including the function $Y$, is 6.28 pb.  If the resummed CSS piece is used
alone, the contribution is reduced to 4.73 pb.  
This answer can be compared to the contribution at LO.  
For the same structure functions, the LO $gg$ cross section for the CDF cuts
is 3.18 pb for the scale choice $Q=M_{\gamma\gamma}$.
Therefore, the effect of including part of the NLO contribution
to the $\gamma\gamma$ process is to approximately double the
LO $gg$ box contribution to the cross section.  
This increase indicates that the exact NLO
correction can be large for the $gg$ subprocess and
motivates a full calculation.

The predictions for the \D0~cuts and a comparison to data are shown
in Figs.~\ref{Fig:D00}--\ref{Fig:D02}.  
Because of the steep distribution in the transverse
momentum of the individual photons, the higher $p_T^\gamma$ threshold
in the \D0~case significantly reduces the total cross section.
Otherwise, the behavior is similar to the resummed calculation
shown for the CDF cuts.  The \D0~data plotted in the figures
are not corrected for experimental resolution.
To compare with the uncorrected \D0~data with the kinematic cuts
$p^{\gamma_1}_T >14$\,GeV, $p^{\gamma_2}_T >13$\,GeV
and $\eta^\gamma<1.0$, an ``equivalent'' set of cuts
is used in the theoretical calculation:
$p^{\gamma_1}_T >14.9$\,GeV, $p^{\gamma_2}_T >13.85$\,GeV,
and $\eta^\gamma<1.0$ \cite{d0www}. The effect of this 
``equivalent'' set is to reduce the theoretical rate in the
small $M_{\gamma \gamma}$ region.

While the agreement in both shapes and absolute rates is generally
good,
there are some discrepancies between the resummed prediction and the data
as presented in these plots.
At small $Q_T$ (Fig.~\ref{Fig:CDF1}) and large $\Delta\phi$ 
(Fig.~\ref{Fig:CDF2}), where the CDF cross section is large, the theoretical
results are beneath the data.  
Since this is the kinematic region in which the nonperturbative 
physics is important, better agreement can be obtained if
the nonperturbative function is altered.
In Fig.~\ref{Fig:D00}, the calculated $M_{\gamma\gamma}$ distribution
 is larger
than the \D0~data at large $M_{\gamma\gamma}$, while
the calculation appears to agree with the CDF data in Fig.~\ref{Fig:CDF0}.
The small discrepancy in Fig.~\ref{Fig:D00} at large values
of $M_{\gamma\gamma}$ is not understood. 
(The systematic errors of the data, which are about
25\% \cite{d0www}, are not included in this plot.)
On the other hand, Figs.~\ref{Fig:D01} and \ref{Fig:D02}
show that the resummed calculation is {\it beneath} the data 
at large $Q_T$ or small $\Delta\phi$.
The discrepancies in Figs.~\ref{Fig:D01} and \ref{Fig:D02} may
result from the approximations
made in the $gg$ process (notice the kinks in the dot--dashed curves).
A complete NLO calculation for the $gg$ subprocess
is needed, and may improve the comparison with data
for small $\Delta\phi$.

Because of the uncertainty in the prediction for the $gg$
contribution of the resummed calculation, the distributions
in $M_{\gamma\gamma}, Q_T$ and $\Delta\phi$
are shown in Figs.~\ref{Fig:qtcut0}--\ref{Fig:qtcut2} 
for the CDF cuts and the additional 
requirement that
$Q_T < M_{\gamma\gamma}$.  This additional requirement
should significantly reduce the theoretical uncertainty for
large $Q_T$ and small $\Delta\phi$.

In Fig.~\ref{Fig:qtcut1}, the lower of the two solid curves in
the $Q_T$ distribution shows
the prediction of the pure NLO ${\cal O}(\alpha_s)$ (fixed--order)
calculation, without resummation, for the $q\bar q$ and $qg$ 
subprocesses, excluding fragmentation.  For $Q_T \grsim 25$ GeV,
the lower solid curve is very close to the long--dashed
($q\bar q+qg$) curve obtained after resummation, as is expected.
As $Q_T$ decreases below $Q_T\simeq 25$ GeV, all--orders
resummation produces significant changes.  Most apparent, perhaps,
is that the $Q_T\to 0$ divergence in the fixed--order calculation is
removed.  However, there is also a marked difference in shape
over the interval $5<Q_T<25$ GeV between the fixed--order
$q\bar q+qg$ result and its resummed counterpart.  
These are general features in a comparison of resummed and NLO
calculations \cite{resum2}--\cite{qqres}.

\subsection{Fixed--Target Energy}

The fixed--target
experiment E706 \cite{e706data} at Fermilab 
has collected diphoton data from the collision of
a $p$ beam on a $Be$ ($A=9.01, Z=4$) target at $\sqrt{S}=31.5$ GeV.
The kinematic cuts
applied to the resummed prediction in the center--of--mass frame
of the beam and target
are  $p_T^\gamma > 3$ GeV and $|\eta^{\gamma}|<0.75$.  
No photon isolation is required.
The same phenomenological inputs are used for
this calculation as for the calculation at collider energies.  
The $Be$ nucleon target
is treated as having an admixture of $4/9.01$ proton and $5.01/9.01$
neutron parton distribution functions.  The $A$ dependence effect
appears to be small in the prompt photon data (the effect is 
parametrized as
$A^\alpha$ and the measured dependence is $\alpha\simeq 1$), so it
is ignored \cite{sorrell}.

Figures \ref{Fig:E7060}--\ref{Fig:E7062} 
show the same distributions discussed previously. 
Because of the kinematic cuts, the relative contribution of
gluon initiated processes is highly suppressed, except at
large $Q_T$, where the $gg$ box contribution is seen to
dominate, and at large $M_{\gamma\gamma}$ where the
$qg$ contribution is dominant.  
The fragmentation contribution (not shown) is
minimal (of a few percent). 
The dominance of $gg$ resummation over the $q\bar q$ resummation
at large $Q_T$ in Fig.~\ref{Fig:E7061} occurs because
it is more likely (enhanced by the ratio $C_A/C_F=9/4$) 
for a gluon to be radiated from a gluon than a quark line.
The exact height of the distribution is sensitive to
the form of the nonperturbative function 
(in the low $Q_T$ region) and to the approximation
made in calculating the NLO corrections 
(of ${\cal O}(\alpha_{em}^2\alpha_s^3)$)
to the hard scattering.
However, since $Q_T < Q$ is satisfied for the set of 
kinematic cuts, the final answer 
with complete NLO corrections
should not differ significantly from the result reported here.

The scale dependence of the calculation
was checked by comparing
with the result obtained with $C_2=C_1/b_0=0.5$, $C_3=b_0$, and $C_4=1$.
The $q {\bar q}$ rate is not sensitive to the scale 
choice, and the $gg$ rate increases by less than about 20\%. 
This insensitivity can be understood as follows. 
For the E706 data, the nonperturbative physics completely dominates the 
$Q_T$ distribution.  The perturbative Sudakov resummation is 
not important over the entire $Q_T$ region, and
the NLO $Y$ piece is sizable only for $Q_T > 3$\,GeV where the event
rate is small.  Since the LO $q \bar q$ rate does depend on $\alpha_s$,
and the LO $gg$ rate is proportional to 
$\alpha_s^2(C_2 M_{\gamma \gamma})$, the $gg$ rate increases for a 
smaller $C_2$ value, but the $q \bar q$ rate remains about the same.
In conclusion, the E706 data can be used to constrain the
nonperturbative functions associated with the $q {\bar q}$ and $gg$ 
hard processes in hadron collisions.

\section{Discussion and Conclusions}

Prompt photon pair production at fixed target and collider energies is of 
interest in its own right as a means of probing the dynamics of strong 
interactions.  The process is of substantial interest also in searches for 
new phenomena, notably the Higgs boson. 

In this paper, a calculation is presented of the production rate
and kinematic distributions of
photon pairs in hadronic collisions.  
This calculation incorporates the full content of the next--to--leading 
order (NLO) contributions from the $q {\bar q}$ and $q g$ initial--state
subprocesses, 
supplemented by resummation of contributions to these subprocesses from 
initial state radiation of soft gluons to all orders in the strong 
coupling strength.  The computation also includes important contributions 
from the $g g$ box diagram.  The $gg$ contributions 
from initial--state multiple soft gluons are resummed to all 
orders, but the NLO contribution,
of ${\cal O}(\alpha_{em}^2\alpha_s^3)$,
to the hard scattering subprocess is handled in an approximate 
fashion.  The approximation should be reliable at relatively small values of 
the pair transverse momentum $Q_T$ as compared to the
invariant mass of the photon pair $M_{\gamma\gamma}$.  
At collider energies, the $g g$ 
contribution is comparable to that of the $q {\bar q}$ and $q g$ 
contributions over a significant part of phase space where
$M_{\gamma\gamma}$ is not large, and its inclusion is 
essential.  
The exact ${\cal O}(\alpha_{em}^2\alpha_s^3)$ corrections
to the $gg$ box diagram should
be calculated to test the validity of the approximations made in this
calculation.
Finally, the calculation also includes 
long--distance fragmentation 
contributions at leading order from the subprocess 
$q g \rightarrow \gamma q$, 
followed by fragmentation of the final quark, $q \rightarrow \gamma X$. 
After photon isolation, fragmentation plays a relatively minor role.  The 
fragmentation contribution is computed in two ways: first, in the standard 
parton model collinear approximation and second, with a Monte Carlo shower 
simulation.  This overall calculation is the most complete treatment 
to date of photon pair production in hadronic collisions.  Resummation plays 
a very important role particularly in the description of the behavior of the 
$Q_T$ distribution at small to moderate values of this variable, where the 
cross section takes on its largest values.  

The resummed calculation is necessary for a reliable prediction of 
kinematic distributions that depend on correlations between
the photons.  It is a significant improvement over fixed--order
NLO calculations that do not include the effects of initial--state
multiple soft--gluon radiation.  Furthermore, even though the 
 hard scattering  $q \bar q$ and $q g$ subprocesses are computed to
the same order in the resummed and fixed--order NLO calculations, the cross
sections from the two calculations can differ after kinematic cuts 
are imposed \cite{wres}.  

The results of the calculation are compared with data from the CDF and 
\D0~collaborations, and the agreement is generally good in both absolute 
normalization and shapes of the distributions in the invariant mass
$M_{\gamma\gamma}$ 
of the diphoton system, the pair transverse momentum $Q_T$, and
 the difference 
in the azimuthal angles $\Delta\phi$.  
Discrepancies with CDF results at the smallest 
values of $Q_T$ and $\Delta \phi$ near $\pi$ 
might originate from the strong dependence on the nonperturbative
functions in this kinematic region.
In comparison with the \D0~data, there is also evidence for disagreement at 
intermediate and small values of $\Delta \phi$.  The region of intermediate 
$\Delta \phi$, where the two photons are not in a 
back--to--back configuration, 
is one in which the full treatment of three body final--state contributions 
of the type $\gamma \gamma j$ are important, with $j = q$ or $g$.  The 
distributions in Figs. 5 and 8 suggest that an exact calculation of the 
NLO contribution associated with the $g g$ initial channel 
would ameliorate the situation and will be necessary to
describe data at future high energy hadron colliders. 

Predictions are also 
presented in the paper for $p N \rightarrow \gamma \gamma X$ 
at the center--of--mass energy 31.5 GeV, 
appropriate for the E706 fixed--target 
experiment at Fermilab.  The large $Q_T$ and small $\Delta\phi$ behavior
of the kinematic distributions is dominated by the resummation of
the $gg$ initial state.
Nonperturbative physics controls the $Q_T$
distribution, and neither the perturbative Sudakov nor the 
regular NLO contribution plays an important role,
except
in the very large $Q_T$ region where the event rate is small.   
For the E706 kinematics, the requirement $Q_T < Q$ is generally
satisfied. Therefore, the approximate $gg$ calculation presented
in this work should be reliable.

In this calculation, the incident partons are assumed to be collinear with 
the incident hadrons.  A recurring question in the 
literature is the extent to which finite 
``intrinsic'' $k_T$ may be required for a 
quantitative description of data \cite{grabbag,e706data}.  
An important related issue is the 
proper theoretical specification of the intrinsic component
\cite{kt_intrinsic}.
In the CSS resummation formalism, this physics is included by
properly parametrizing the nonperturbative function
${\widetilde W}^{NP}(b)$, which can be measured in
Drell--Yan, $W$, and $Z$ production.
Because photons participate directly in the hard scattering, because their 
momenta can be measured with greater precision than that of hadronic jets 
or heavy quarks, and because the $\gamma \gamma$  final state is a color 
singlet, the reaction $p {\bar p} \rightarrow \gamma \gamma X$ may serve 
as a particularly attractive laboratory for the understanding of the role of 
intrinsic transverse momentum.  
The agreement with data on the $Q_T$ 
distributions in Figs.~\ref{Fig:CDF1} and \ref{Fig:D01} 
is suggestive that the CSS formalism is adequate.
However, the 
separate roles of gluon resummation and the assumed nonperturbative 
function in the successful description of the $Q_T$ distributions are not 
disentangled.  
In the non--perturbative function of Eq.~(\ref{Eq:WNP}), the dependence 
on $b$ 
(and, thus, the behavior of $d \sigma /dQ_T$ at small $Q_T$) is predicted 
to change with both $Q$ and the values of the parton momentum 
fractions $x_i$.  
At fixed $Q$, dependence on the values of the $x_i$ translates into dependence 
on the overall center--of--mass energy of the reaction.  As data with greater 
statistics become available, it should be possible to verify these 
expectations.  In combination with similar studies with data on massive 
lepton--pair production (the Drell--Yan process), it will be possible to 
determine whether the same non--perturbative function is applicable in the 
two cases, as is assumed in this paper.   

The diphoton data may allow a study of the nonperturbative
as well as the perturbative physics associated with 
multiple gluon radiation from the {\it gluon}--initiated hard processes, 
which cannot
be accessed from Drell--Yan, $W^\pm$, and $Z$ data.
With this knowledge, it may be possible to improve
calculations of single photon production and other reactions
sensitive to gluon--initiated subprocesses.
In the \D0~data analysis \cite{d0data}, 
an asymmetric cut is applied on the transverse momenta
($p_T^{\gamma}$) of the two photons in the diphoton event.
This cut reduces the effect of
multiple gluon radiation in the event.
To make the best use of the data for probing the interesting 
multiple gluon dynamics predicted by the QCD theory, a symmetric
$p_T^{\gamma}$ cut should be applied.

\section*{ Acknowledgments}

We thank J. Huston for explaining the Tevatron and fixed--target data.
Also, C.-P.Y. and C.B. thank the CTEQ collaboration and C. Schmidt
for many invaluable discussions, and
S.M. thanks R. Blair,
S. Kuhlmann, J. Womersley,
L. Gordon, and C. Coriano for useful conversations.
This work was supported in part by the NSF under grant PHY--9507683 and
the U.S. Dept. of Energy under grant W--31--109--ENG--38.

\vspace{0.4cm}

\indent


%
%
\begin{figure}[!ht]
\ifx\nopictures Y \else{
\centerline{\ \psfig{figure=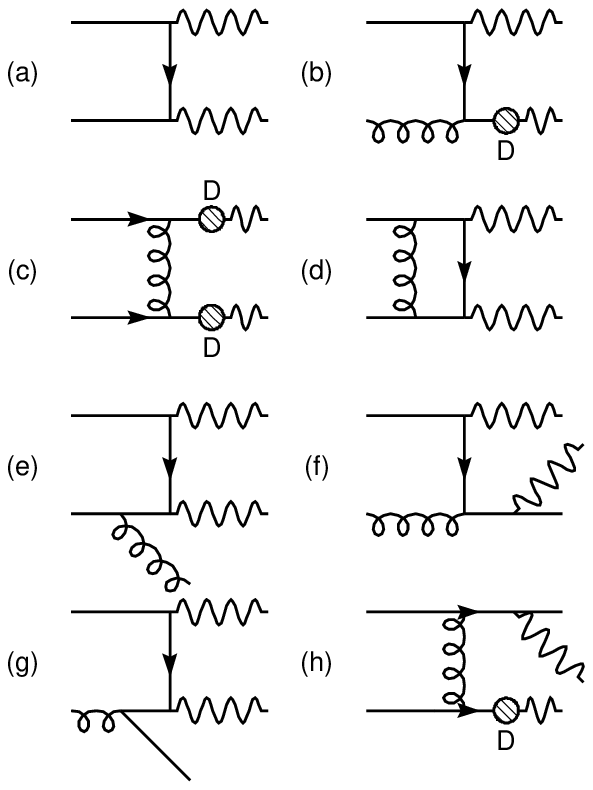,height=10cm} }
} \fi
\caption{
Feynman diagrams representing the leading order and
next--to--leading order contributions to photon pair
production in hadron collisions.  The shaded circles signify
the production of
long--distance fragmentation photons, which are described
by the fragmentation function $D_{\gamma\leftarrow q}$.
}
\label{Fig:Feynman}
\end{figure}
\begin{figure}[!ht]
\ifx\nopictures Y \else{
\centerline{\ \hbox{
\hbox{\psfig{figure=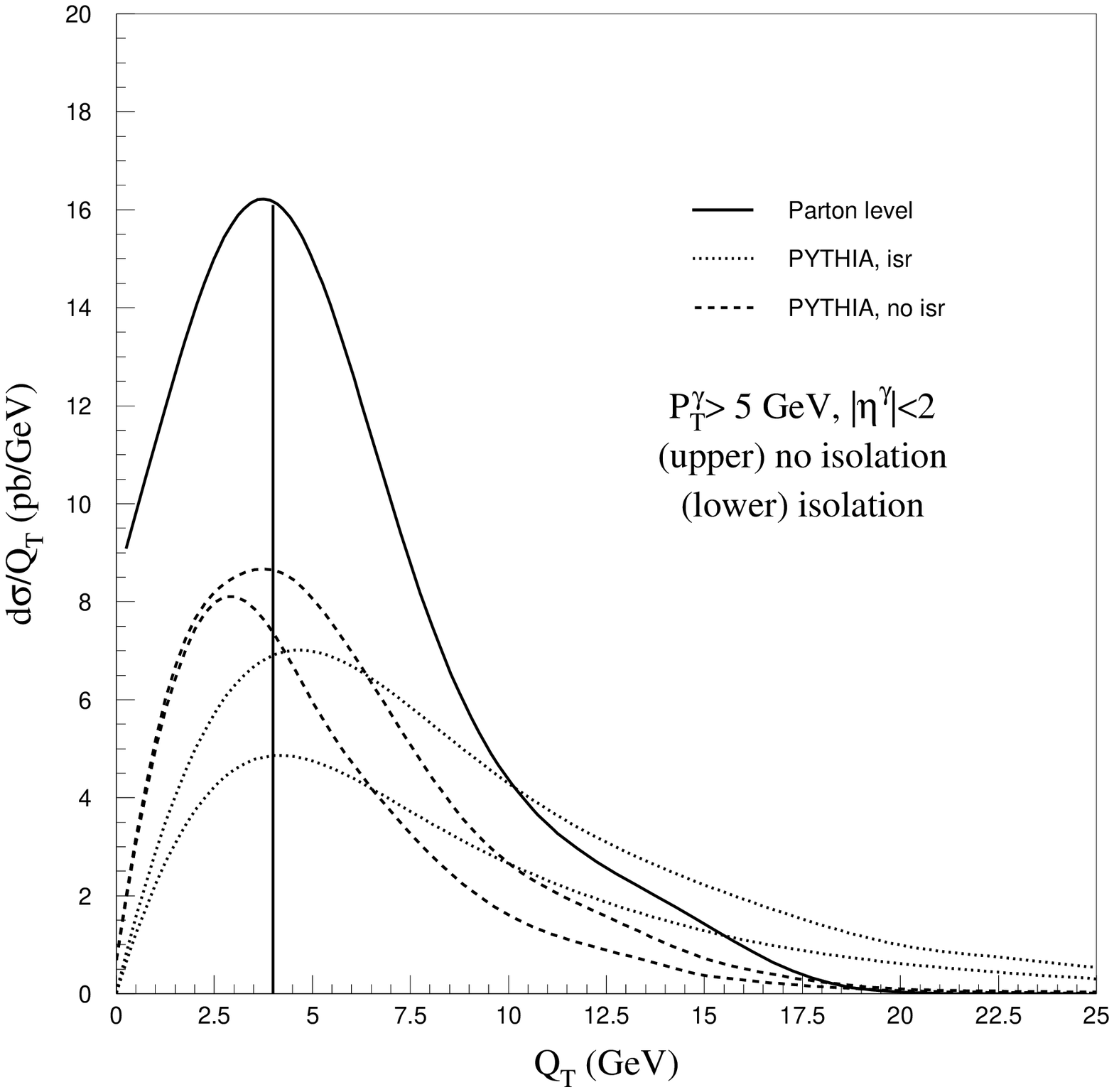,height=7.5cm}}
\hbox{\psfig{figure=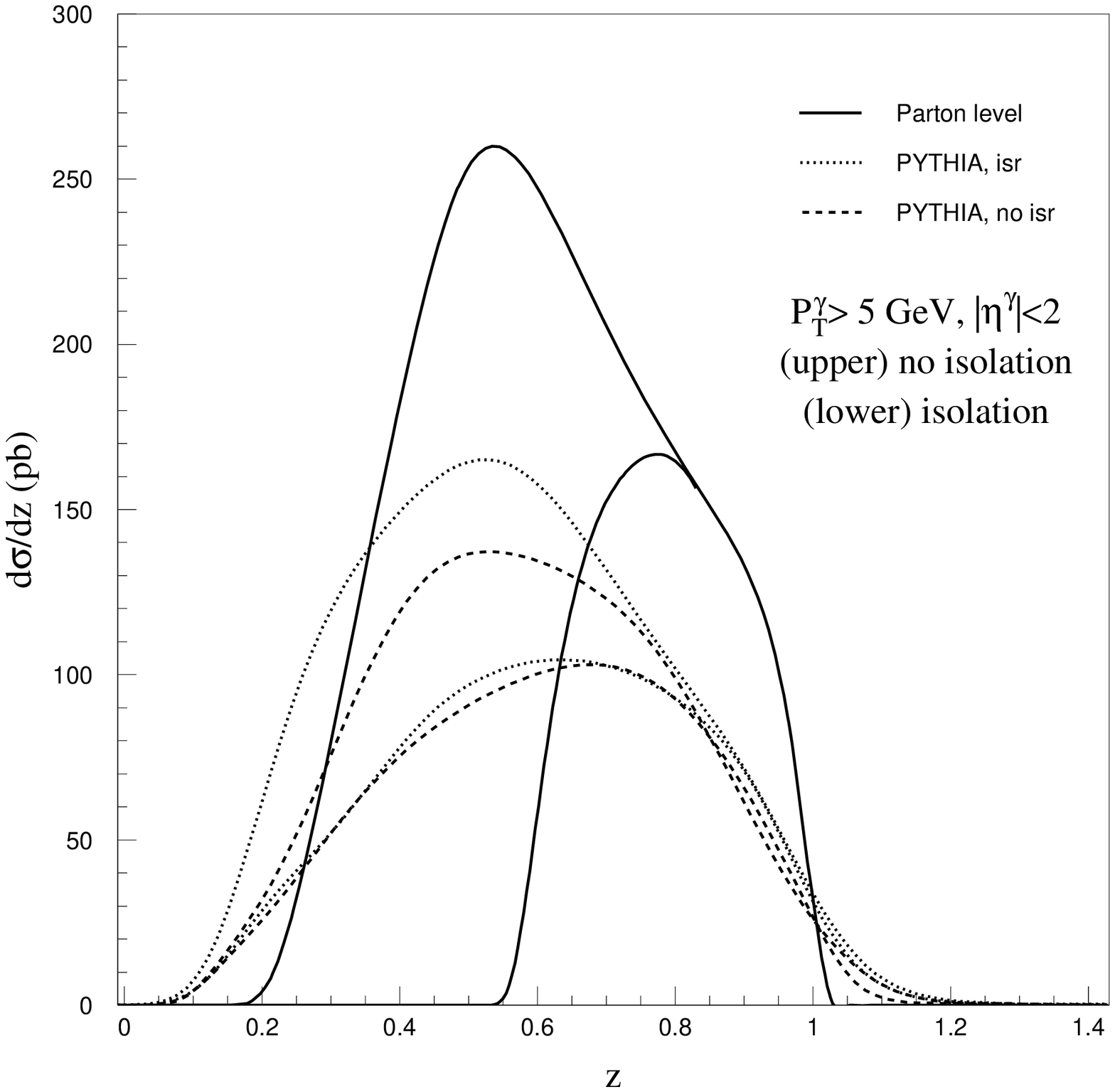,height=7.5cm}} }
}} \fi
\caption{
Comparison of the parton--level and Monte Carlo
fragmentation contributions at the Tevatron.
The upper and lower curves of the same type show
the contribution before and after an isolation
cut.  The left figure shows the transverse momentum
of the photon pair $Q_T$.  The right figure shows the
light--cone momentum fraction carried by the fragmentation
photon.
}
\label{Fig:Frag}
\end{figure}
\begin{figure}[!ht]
\ifx\nopictures Y \else{
\centerline{\psfig{figure=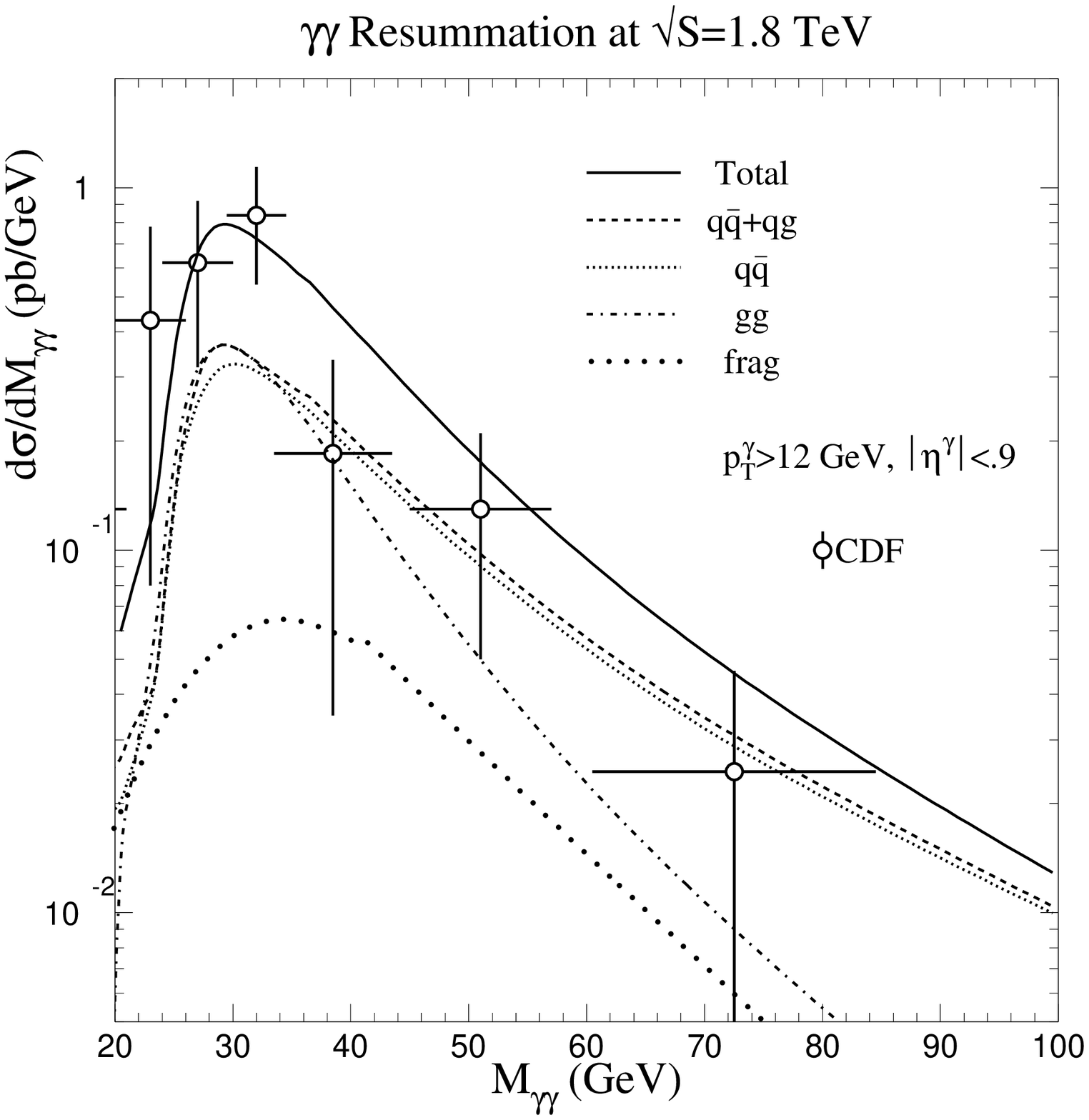,height=10cm} }
} \fi
\caption{The predicted distribution for the invariant mass of the photon
pair $M_{\gamma\gamma}$ from the resummed calculation compared to the
CDF data, with the CDF cuts imposed in the calculation.}
\label{Fig:CDF0}
\end{figure}
\begin{figure}[!ht]
\ifx\nopictures Y \else{
\centerline{\psfig{figure=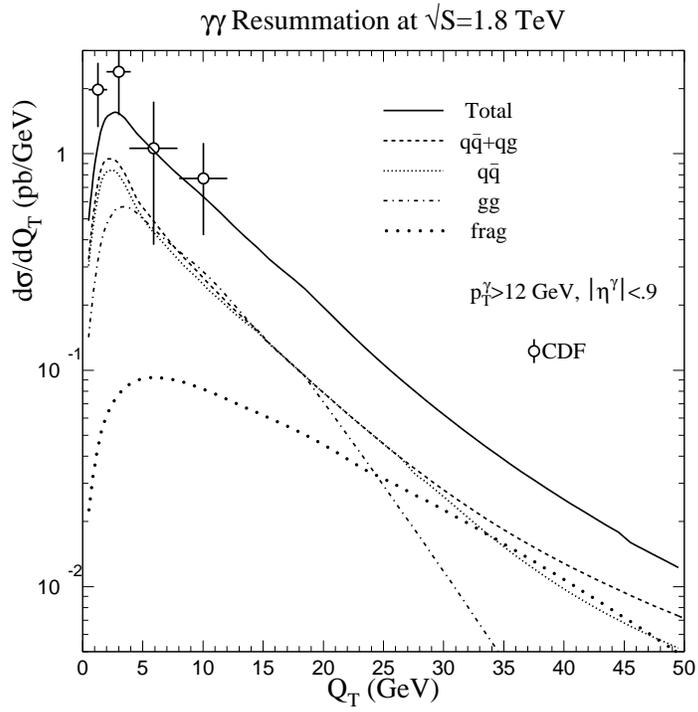,height=10cm} }
} \fi
\caption{The predicted distribution for the transverse momentum of the
photon pair $Q_{T}$ from the resummed calculation compared to the
CDF data, with the CDF cuts imposed in the calculation.}
\label{Fig:CDF1}
\end{figure}
\begin{figure}[!ht]
\ifx\nopictures Y \else{
\centerline{\psfig{figure=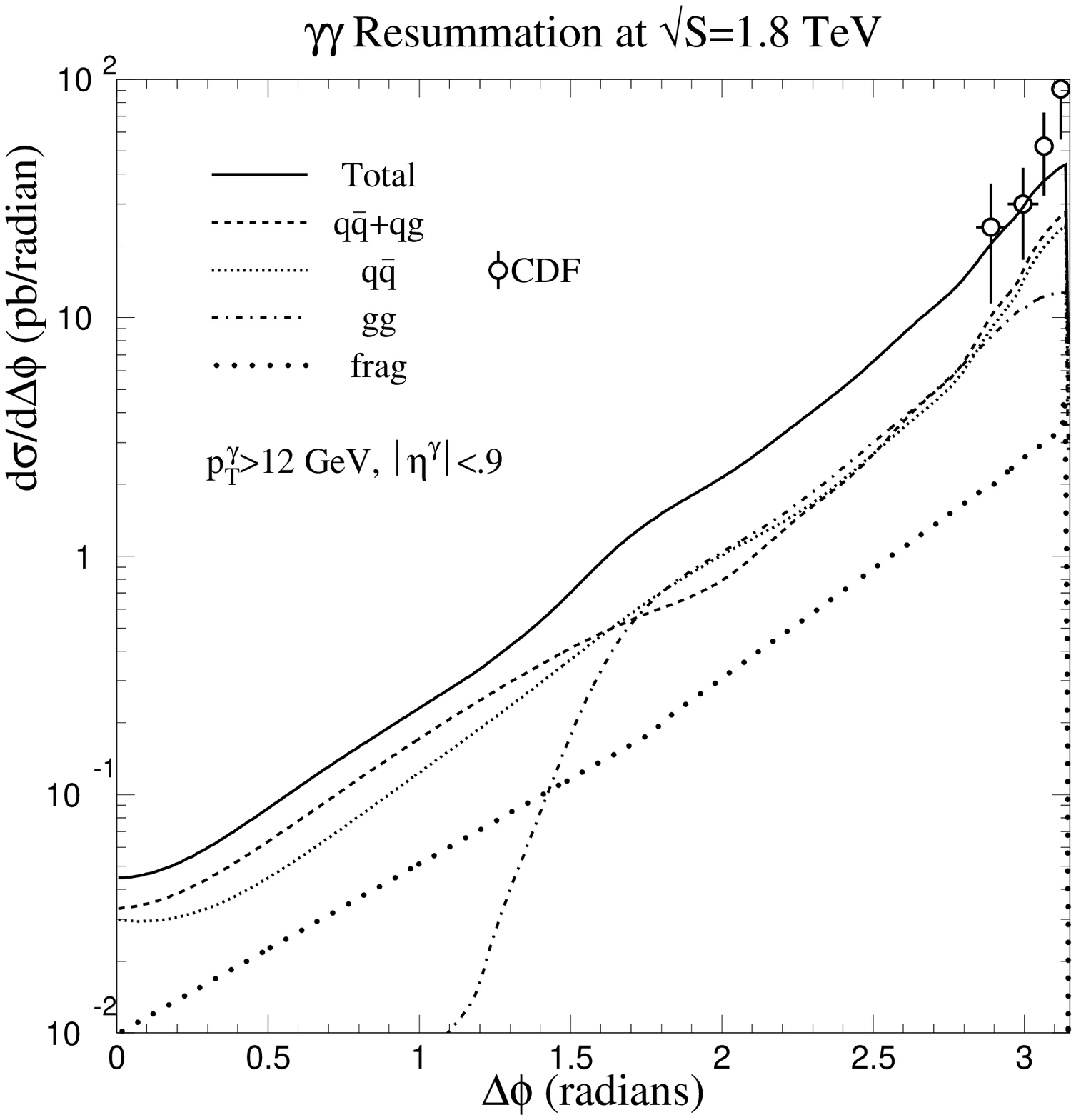,height=10cm} }
} \fi
\caption{The predicted distribution for the difference between the
azimuthal angles of the photons $\Delta\phi_{\gamma\gamma}$ from
the resummed calculation compared to the
CDF data, with the CDF cuts imposed in the calculation.}
\label{Fig:CDF2}
\end{figure}
\begin{figure}[!ht]
\ifx\nopictures Y \else{
\centerline{\psfig{figure=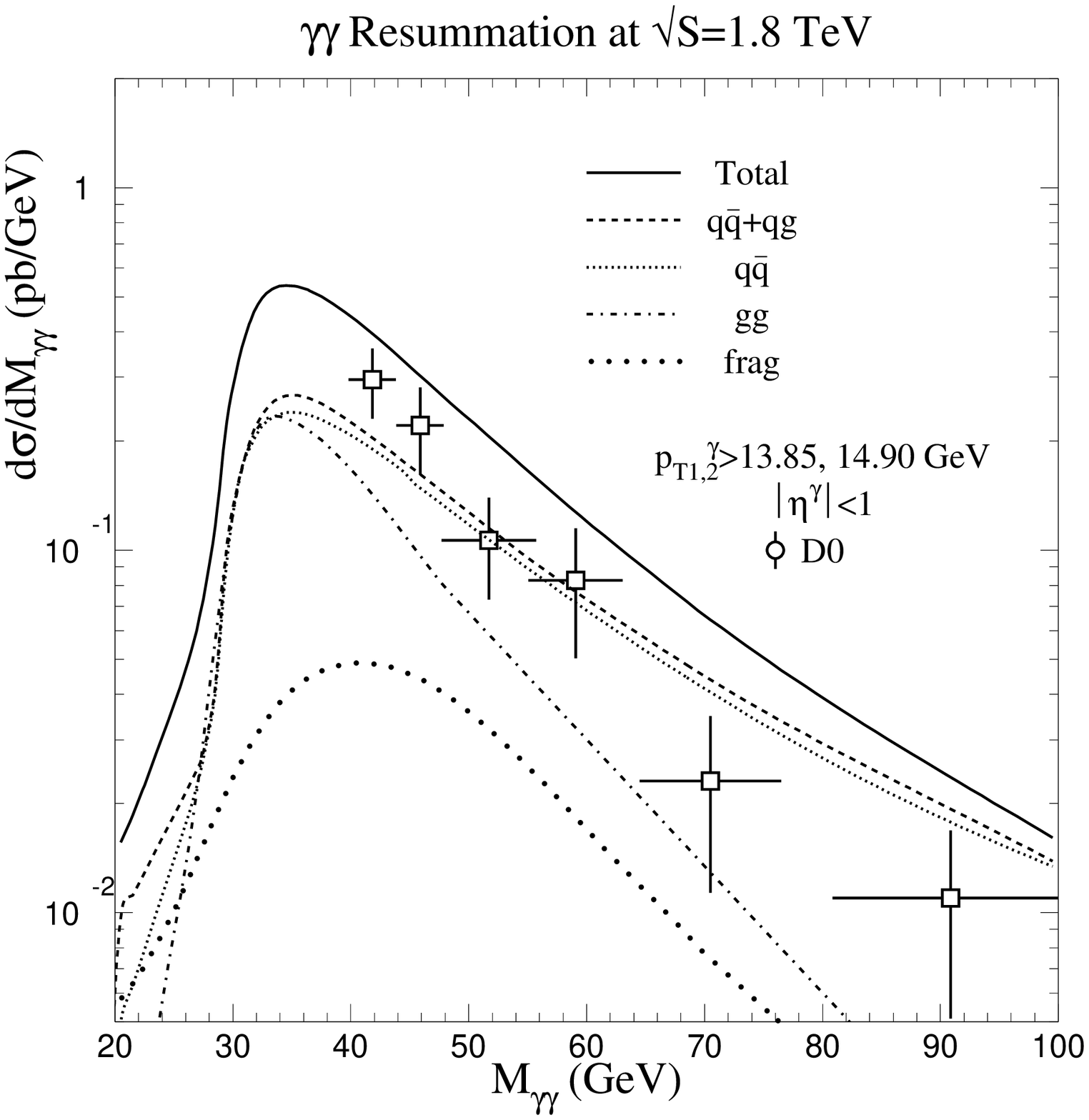,height=10cm} }
} \fi
\caption{The predicted distribution for the invariant mass of the photon
pair $M_{\gamma\gamma}$ from the resummed calculation compared to
the \D0~data, with the \D0~cuts imposed in the calculation.}
\label{Fig:D00}
\end{figure}
\begin{figure}[!ht]
\ifx\nopictures Y \else{
\centerline{\psfig{figure=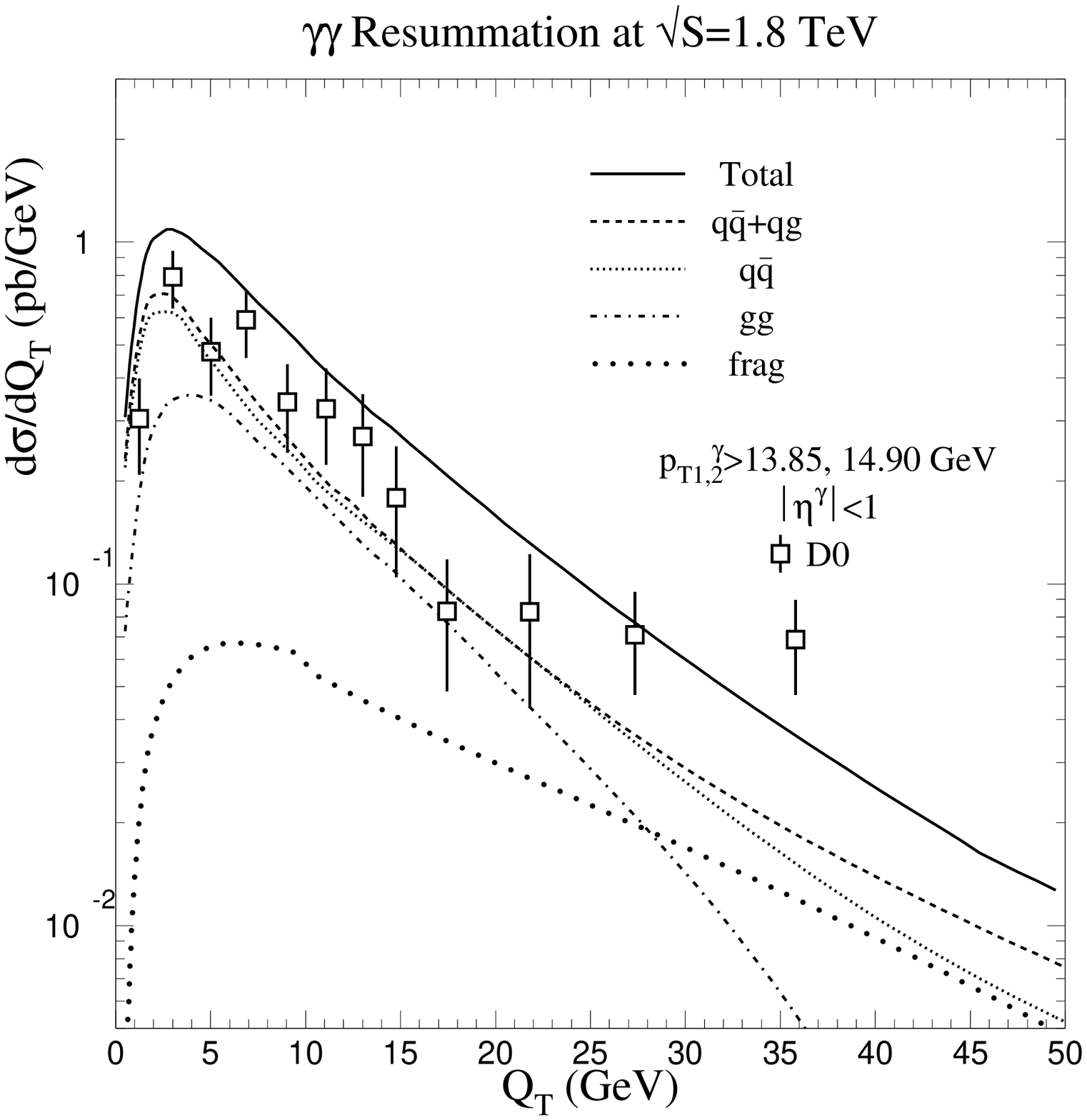,height=10cm} }
} \fi
\caption{The predicted distribution for the transverse momentum of the
photon pair $Q_{T}$ from the resummed calculation compared to
the \D0~data, with the \D0~cuts imposed in the calculation.}
\label{Fig:D01}
\end{figure}
\begin{figure}[!ht]
\ifx\nopictures Y \else{
\centerline{\psfig{figure=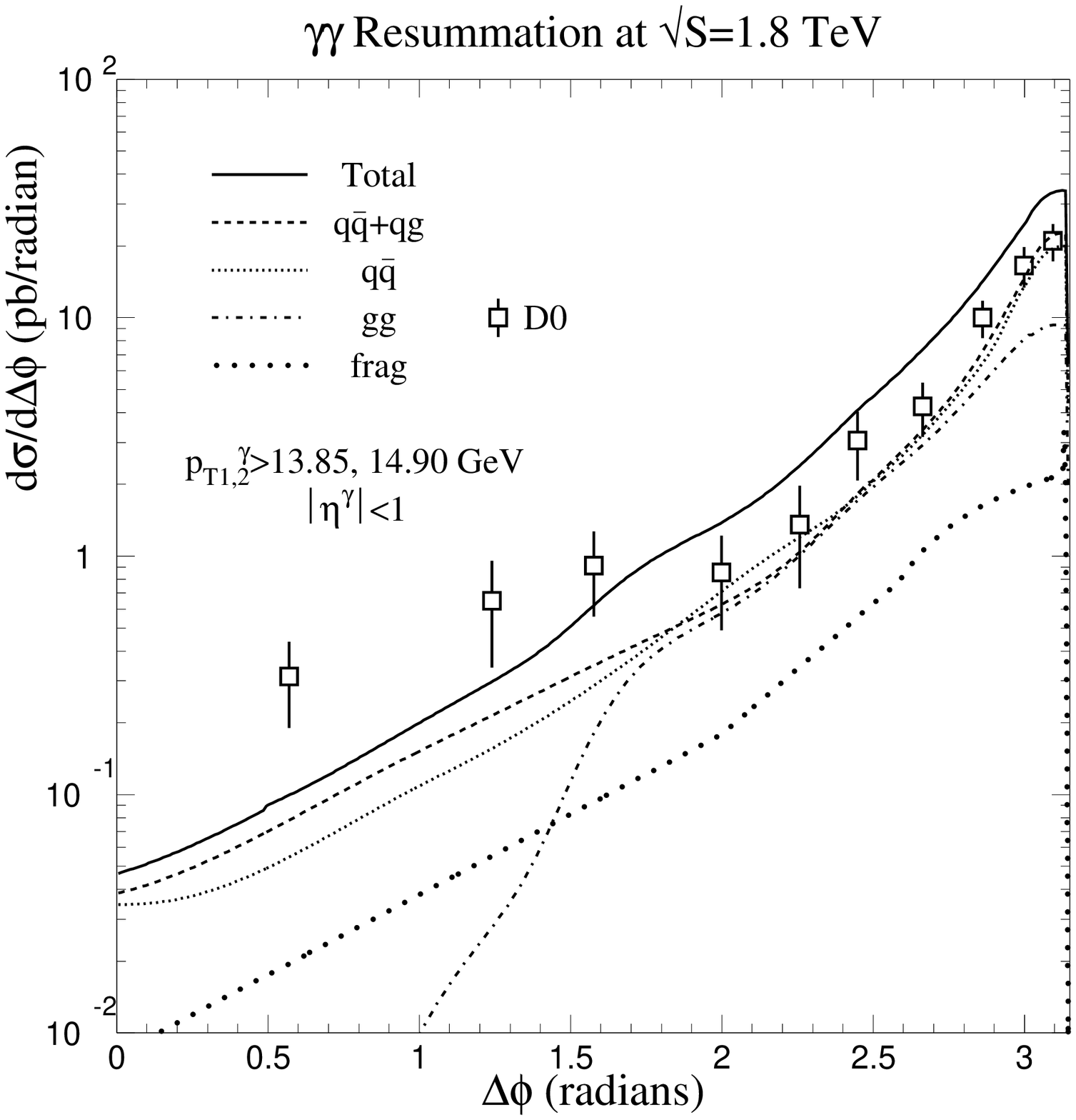,height=10cm} }
} \fi
\caption{The predicted distribution for the difference between the
azimuthal angles of the photons $\Delta\phi_{\gamma\gamma}$ from
the resummed calculation compared to 
the \D0~data, with the \D0~cuts imposed in the calculation.}
\label{Fig:D02}
\end{figure}
\begin{figure}[!ht]
\ifx\nopictures Y \else{
\centerline{\psfig{figure=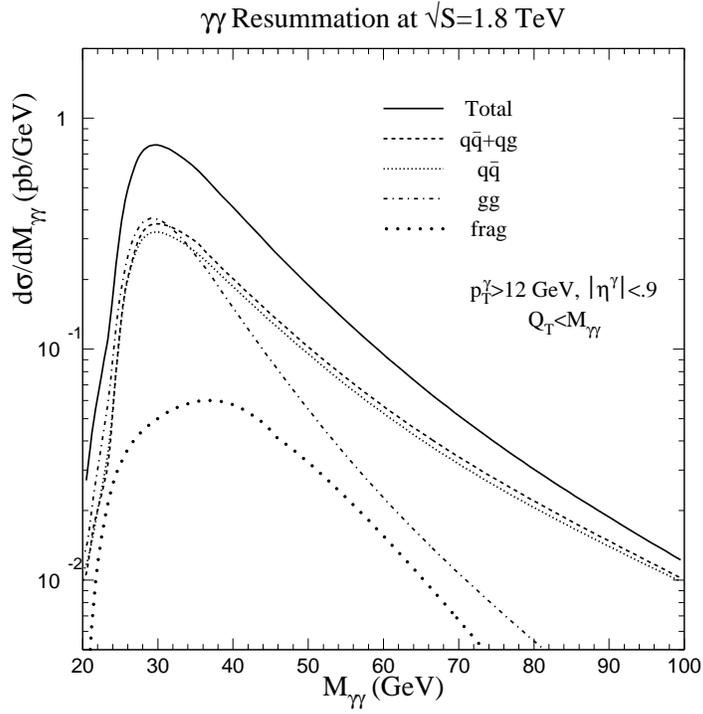,height=10cm} }
} \fi
\caption{The predicted distribution for the invariant mass of the photon
pair $M_{\gamma\gamma}$ from the resummed calculation.
The additional cut $Q_T < M_{\gamma\gamma}$ has been applied to reduce
the theoretical uncertainty.}
\label{Fig:qtcut0}
\end{figure}
\begin{figure}[!ht]
\ifx\nopictures Y \else{
\centerline{\psfig{figure=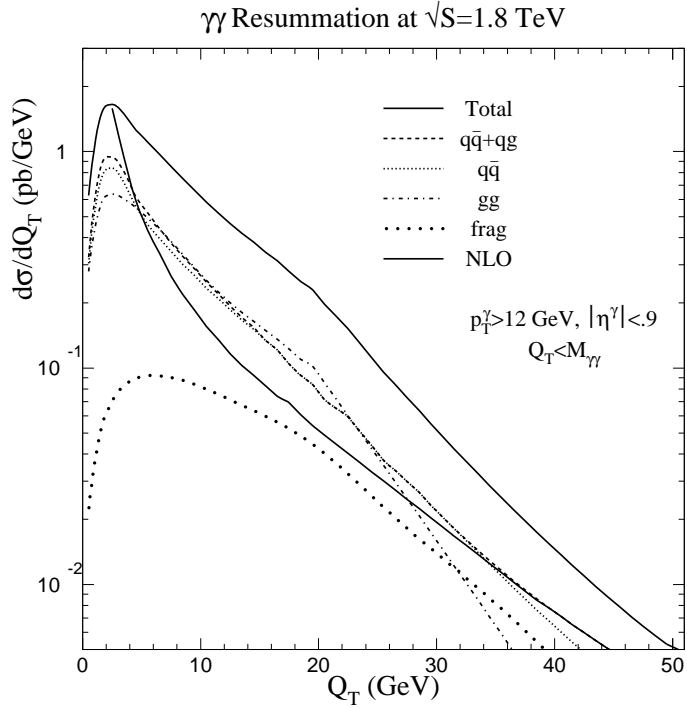,height=10cm} }
} \fi
\caption{The predicted distribution for the transverse momentum of the
photon pair $Q_{T}$ from the resummed calculation.
The additional cut $Q_T < M_{\gamma\gamma}$ has been applied to reduce
the theoretical uncertainty.
The lower solid curve shows the prediction of the pure NLO
(fixed--order) calculation for the $q\bar q$ and $qg$ subprocesses,
but without fragmentation contributions.}
\label{Fig:qtcut1}
\end{figure}
\begin{figure}[!ht]
\ifx\nopictures Y \else{
\centerline{\psfig{figure=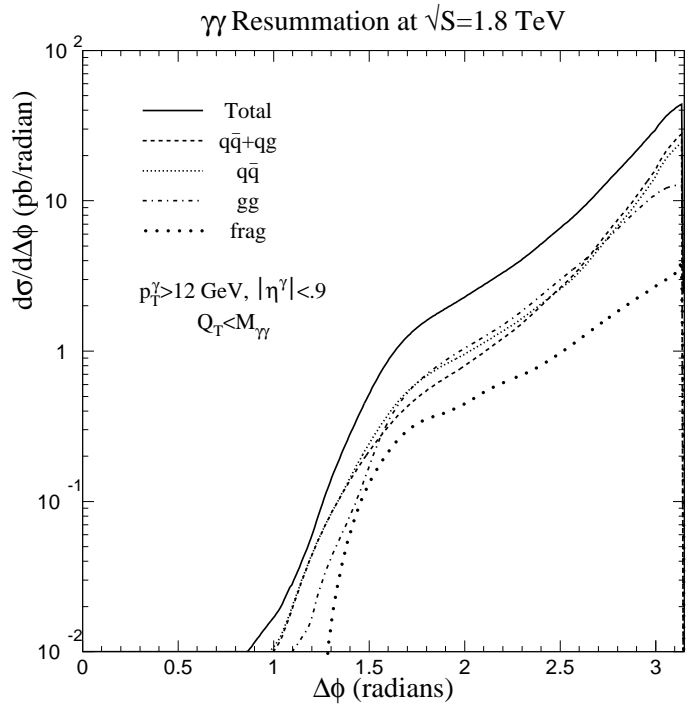,height=10cm} }
} \fi
\caption{The predicted distribution for the difference between the
azimuthal angles of the photons $\Delta\phi_{\gamma\gamma}$ from
the resummed calculation.
The additional cut $Q_T < M_{\gamma\gamma}$ has been applied to reduce
the theoretical uncertainty.}
\label{Fig:qtcut2}
\end{figure}
\begin{figure}[!ht]
\ifx\nopictures Y \else{
\centerline{\psfig{figure=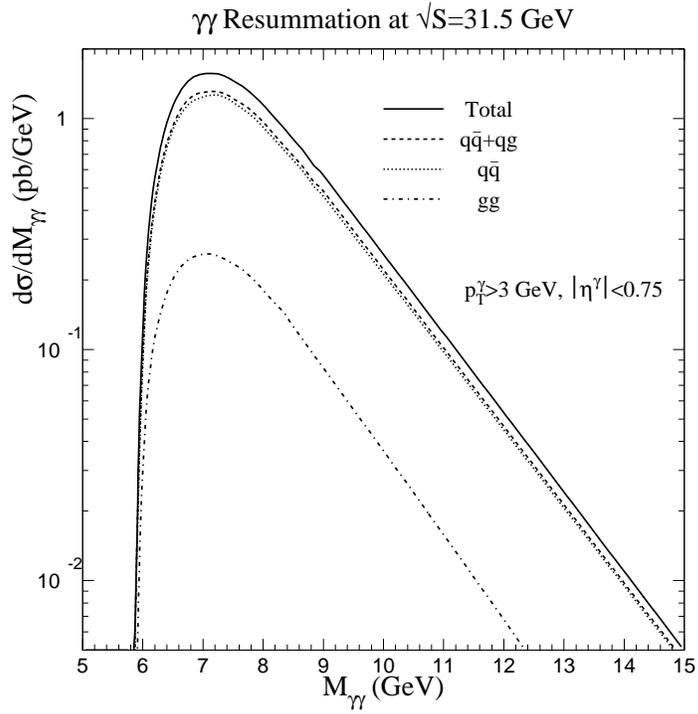,height=10cm} }
} \fi
\caption{The predicted distribution for the invariant mass of the photon
pair $M_{\gamma\gamma}$ from the resummed calculation 
appropriate for $pN\to\gamma\gamma X$ at $\protect{\sqrt{S}}$=31.5 GeV.}
\label{Fig:E7060}
\end{figure}
\begin{figure}[!ht]
\ifx\nopictures Y \else{
\centerline{\psfig{figure=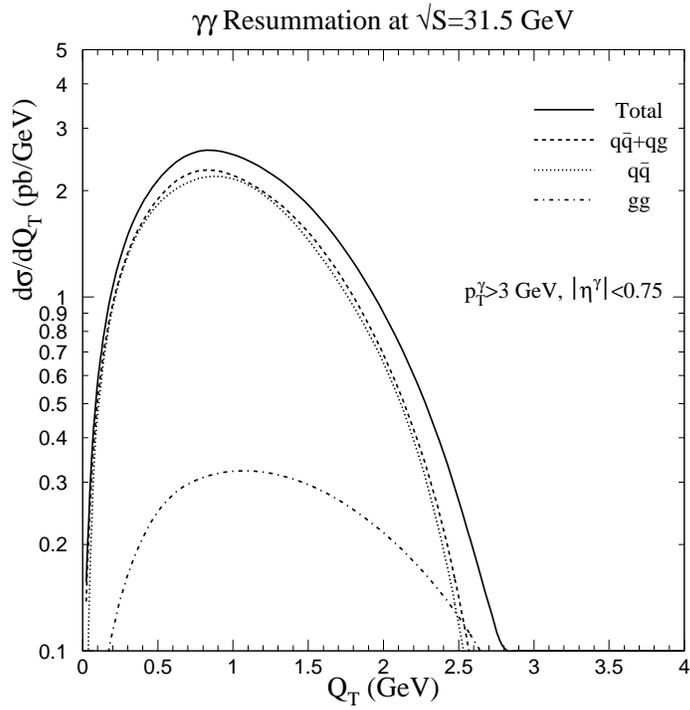,height=10cm} }
} \fi
\caption{The predicted distribution for the transverse momentum of the
photon pair $Q_{T}$ from the resummed calculation 
appropriate for $pN\to\gamma\gamma X$ at $\protect{\sqrt{S}}$=31.5 GeV.}
\label{Fig:E7061}
\end{figure}
\begin{figure}[!ht]
\ifx\nopictures Y \else{
\centerline{\psfig{figure=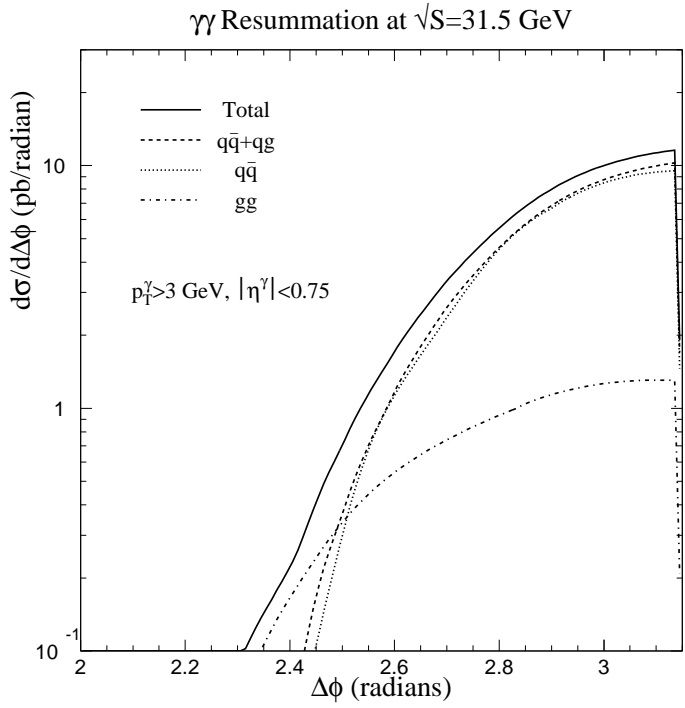,height=10cm} }
} \fi
\caption{The predicted distribution for the difference between the
azimuthal angles of the photons $\Delta\phi_{\gamma\gamma}$ from
the resummed calculation 
appropriate for $pN\to\gamma\gamma X$ at $\protect{\sqrt{S}}$=31.5 GeV.}
\label{Fig:E7062}
\end{figure}
\end{document}